\documentclass[twocolumn,preprintnumbers,superscriptaddress]{revtex4-2}
\usepackage{hyperref}
\usepackage[utf8]{inputenc}

\usepackage{soul}
\usepackage{url}
\usepackage{mathtools}
\usepackage{braket}
\usepackage{amsmath}
\usepackage{dsfont}
\usepackage[english]{babel}
\usepackage{graphicx} 
\usepackage{tikz}
\usetikzlibrary{quantikz}

\usepackage{hyperref}
\usepackage{natbib} 
\usepackage{physics}

\usepackage{listings}
\usepackage{xcolor}
\usepackage{amssymb}

\newcommand{\LDIG}{
Lighthouse Disruptive Innovation Group, LLC
7 Broadway Terrace, Apt 1
Cambridge MA 02139
Middlesex County, Massachusetts (USA)
}

\newcommand{\DSD}{
Engineering Department
Research Group on Data Science for the Digital Society
La Salle - Universitat Ramon Llull
Carrer de Sant Joan de La Salle, 42
08022 Barcelona (Spain)
}
\newcommand{\UVA}{
Universidad de Valladolid
C/Plaza de Santa Cruz, 8, 
47002 Valladolid (Spain)
}

\graphicspath{{./images/}}

\begin{document}
\title{GPS: A new TSP formulation for its generalizations type QUBO}

\author{Saúl González Bermejo}
\affiliation{\UVA}
\email{saul.gonzalez.bermejo@alumnos.uva.es}

\author{Guillermo Alonso-Linaje}
\affiliation{\UVA}
\email{guillermo@alonso-linaje.com}

\author{Parfait Atchade-Adelomou}
\affiliation{\DSD}
\email{parfait.atchade@salle.url.edu}
\affiliation{\LDIG}
\email{parfait.atchade@lighthouse-dig.com}
\date{Oct 2021}

\begin{abstract}
We propose a new Quadratic Unconstrained Binary Optimization (QUBO) formulation of the Travelling Salesman Problem (TSP), with which we overcame the best formulation of the Vehicle Routing Problem (VRP) in terms of the minimum number of necessary variables. After, we present a detailed study of the constraints subject to the new TSP model and benchmark it with MTZ and native formulations. Finally, we tested whether the correctness of the formulation by entering it into a QUBO problem solver. The solver chosen is a D-Wave\_2000Q6 quantum computer simulator due to the connection between Quantum Annealing and QUBO formulations.


\textbf{KeyWords:} Quantum Computing, Quantum Annealing, Combinatorial Optimization, QUBO, TSP, VRP
\end{abstract}

\maketitle

\section{Introduction}\label{sec:intro}
%
The Travelling Salesman Problem, known as TSP \cite{gavish1978travelling}, is one of the most studied statements belonging to the combinatorial optimization problems. In this, we are given a set of cities and the distances between them with which we try to find the best route to travel all the towns, minimizing the total length.

Both the TSP and its more well-known derivative, the Vehicle Routing Problem (VRP), are routing problems with a great impact on most of the issues in our society. For this reason, and because both are NP-Hard \cite{hochba1997approximation}, the scientific community has not stopped looking for a better formulation that makes their resolution efficient. But, unfortunately, we cannot use traditional search methods based on differentiability when defining the problem with discrete variables.

One of the models that allows us to write TSP like problems more generically is Quadratic Unconstrained Binary Optimization (QUBO) \cite{lewis2017quadratic}. QUBO is a framework that enables us to model problems in a quadratic form subject to linear restrictions natively. However, with the help of penalty functions, it is possible to reformulate the tasks of order greater than two and inequality constraints to the QUBO model.
Another characteristic that makes QUBO a very important modelling environment is its close connection with the Ising model \cite{cipra1987introduction}. The QUBO model constitutes a central problem for adiabatic quantum computing \cite{McGeoch2013}, which is solved through a physical process called quantum annealing \cite{mcgeoch2014adiabatic,brooke1999quantum}.

It is known that the best current QUBO formulation of the TSP requires $ N^2 $ binary variables \eqref{sec:TSPn2}). However, when we try to generalize this formulation to some set of problems such as VRP, we find that polynomial terms of order greater than $ 2 $ appear in these models. As QUBO modelling requires that the function to minimize must be quadratic, it is necessary to decrease the degree of these terms by introducing auxiliary variables, which greatly increases the number of required variables. This is crucial to achieving good results through the solvers dedicated to it, and especially if it is going to be implemented in a quantum computer. \\

Quantum annealing is the paradigm of using quantum processes to solve combinatorial optimization problems. This paradigm uses entropy as a target to force exploration, given that any function that smoothes the probability in the search space can have the same purpose according to the adiabatic theorem \cite{Born1928, farhi2000quantum}. \\ The D-Wave Quantum Processor Unit (QPU) is considered as a heuristic that minimizes the objective QUBO functions using a physically performed version of quantum annealing.

This shows how the number of variables in the QUBO model is related to the number of qubits in a quantum computer \cite{bian2010ising}. 


The VRP encompasses two different problems: one in which the distance travelled by vehicles subject to capacity restrictions is minimized \cite{toth2002vehicle, toth2002models, ralphs2003capacitated, a14070194} and another, in which the time spent it takes cars to complete their routes is minimized. In this article, everything related to the VRP will optimize the time to complete these routes, equivalent to reducing the total of the distances travelled by all vehicles.

As we will explain later in the section dedicated to the VRP, to generalize a QUBO formulation from the TSP to the VRP, the objective function for calculating the distance travelled by the vehicles must be linear so that the formulation discussed above with $ N^2$ variables cannot be used.

The most used QUBO model of the TSP that can represent distance linearly use more than $ N ^ 3 $ variables (native TSP formulation), although there is indeed a formulation that uses $ N ^ 2log_2 (N) $ variables; this formulation is known as MTZ  \cite{miller1960integer}. But generalizing the MTZ formulation for the VRPs that minimizes the distances travelled by all the vehicles to the maximum give us quadratic restrictions and, therefore, a penalty function of the order greater than 2.

Our purpose in this work is to present a new QUBO model of the TSP whose the travelled distance's calculation is linear, and that uses only $ 3N ^ 2 $ variables, considerably improving the existing TSP models (both of $ N ^ 3 $ and $ N ^ 2 log_2 (N) $ variables). Furthermore, this new formulation of the TSP that we call GPS will be generalized to define an efficient formulation considering the number of variables of a new VRP formulation. Unfortunately, after reviewing state of the art, we have not found a QUBO formulation of the VRP that minimizes the maximum of the distances travelled by all the vehicles, so we have not been able to make comparisons.  When we talk about the number of variables in a model, we will describe it according to its dominant term, so for a model that, for example, requires $ N ^ 3 + 3N ^ 2 + 2N $ variables, we will say that it is modelling with $ N ^ 3 $ variables since it gives us enough information on its scalability.

The document is organized as follows. Section \eqref{sec:motivation} presents our main motivation behind this work. Section \eqref{sec:relatedwork} shows previous work on the TSP algorithm and its derivatives. Section \eqref{sec:AQC} will present the QUBO framework and its connection to quantum annealing. In the section \eqref{sec:modelTSP}, we recall the native formulation of TSP and the MTZ QUBO model. Section \eqref{sec:modelGPS} presents our TSP proposal with the improvements in the numbers of variables. A generalization of our contribution is seen in Section \eqref{sec:modelVRP} where we propose our VRP into the QUBO model. Section \eqref{sec:resultados} present the obtained results, and finally, Section \eqref{sec:conclusions} concludes the work carried out, and we open ourselves to some lines of the future of the proposed model.


\section{Motivation}\label{sec:motivation}

Our main motivation is to find a suitable formulation that uses the minimum number of variables; therefore, the minimum number of qubits when implementing said models in quantum computers.
This motivation is boosted by solving the problem presented in our article \cite{a14070194} when we desire that the mobile robots minimize the time, which is equivalent to reducing the maximum of the distances travelled by all the vehicles. This implies reducing the number of qubits necessary to implement this model in this era of very few qubits.

\section{Related Work}\label{sec:relatedwork}

In the mid-1920s, these two references, \cite{boruvka1926minimal, boruvka1926prispevek} were the first articles to provide a solution to the minimal spanning tree (MST) problem. Based on these works, the mathematical researcher, Joseph B. Kruskal Jr, applied these solutions to the TSP \cite{kruskal1956shortest}, giving life to some of the first solutions to this problem that will arise during the next decades.

Almost at the end of the sixties, this work \cite{Bellmore1968} offered a compilation and synthesis of the research on the travelling salesman problem. Its authors began by defining the problem and presenting several relevant theorems. They also classified and detailed the solution techniques and computational results. Before that, in the mid-1960s, the TSP started to emerge in many different contexts. This article \cite{Lenstra1975} highlights some applications that began to gain space in everyday life, such as vehicle routing or job shop scheduling problems. Other applications such as planning, logistics and the manufacture of electronic circuits became of particular interest.

By making a few small modifications to the original TSP, we could apply it in many fields such as SWP \cite{adelomou2020formulation} and DNA sequencing \cite{Lee2004, Ball2000} among others. In this last application, the concept of `city' would come to be fragments of DNA and the idea of `distance', a measure of similarity between the pieces of DNA. In many applications, additional restrictions such as resource limits or time windows make the problem considerably difficult.

Computationally, the TSP \cite{pataki2003teaching} is an NP-Hard problem within combinatorial optimization. As an NP-Hard problem, it is computationally complex, and heuristics are continually being developed to get as close as possible to the optimal solution. However, considering the computational complexity nature of these problems, the new approach that quantum computing takes is very promising.

Many works are related to the standard/native TSP or some related variant in a quantum environment within this new approach. For example, in this work \cite{martovnak2004quantum}, the authors introduced a different quantum annealing scheme based on a path-integral Monte Carlo processes to address the symmetric version of the Travelling Salesman Problem (sTSP). In these other articles \cite{warren2013adapting, warren2017small}, the authors did a comparative study using the D-Wave platform to evaluate and compare the efficiency of quantum annealing with classical methods for solving standard TSP.

In this reference, \cite{greco2008traveling} several comparisons of heuristic techniques were made for some TSP Libraries (TSPLIB) \cite{reinelt1991tsplib} problems, both symmetric and asymmetric, and their results have been compared to other methods such as Self Organizing Maps and Simulated Annealing \cite{crosson2016simulated}.
In both cases, the local search technique was applied to the results found with Wang's Recurrent Neural Network with "Winner Takes All" that improved the Self Organizing Maps \cite{leung2004expanding}.
Other techniques such as the co-adaptive neural network approach to the Euclidean
Travelling Salesman Problem \cite{cochrane2003co} are important too.

One of the generalizations of the TSP, known as the VRP, was studied on the D-Wave platform \cite{feld2019hybrid, irie2019quantum}. In tasks where routing and planning capacity (time) is required, the TSP with time windows (TSPTW) was generalized \cite {focacci2002hybrid, edelkamp2013algorithm}, and has high inherent complexity and presents enormous resolution difficulties. In these references \cite{adelomou2020formulation, AtchadeAdelomou2020, adelomou2020using,atchadeadelomou2021quantum}, the authors modelled combinatorial optimization problems in which social workers visit their patients at their respective homes and attend to them at a specific time, called Social Workers' Problem (SWP). SWP is a significant problem because additional time constraints allow more realistic scenarios to be modelled than native TSP. The optimal or near-optimal solution for such issues is important in minimizing distance and time and environmental problems such as reducing fuel consumption.

The generalization of the TSP that we will use in work will be the VRP. However, there are other TSP derivatives, such as the Job Shop Scheduling Problem (JSSP) \cite {applegate1991computational} that are not included in the study of this work.

During state of the art carried out, we have found several articles \cite{papalitsas2019qubo, feld2019hybrid, irie2019quantum} that solve the TSP and VRP (focusing on minimizing distance and not time) for annealing computers \cite{brooke1999quantum,boixo2014evidence,crosson2016simulated}. However, the number of variables is still intractable for the current size of quantum computers. For this reason, we propose a new TSP formulation with a representation of the linear distance that uses only $ 3N ^ 2 $ variables, which we will use to outperform the current best VRP modelling in terms of the number of required variables. \\
For example, a possible formulation of the VRP uses $ N ^ 3 $ variables where $ N $ is the number of cities, so with only $10$ towns, we would go to 1000 necessary variables. In quantum computing, each of these variables can be represented with a qubit, and that is why 1000 qubits computers would be needed to carry out these tasks. However, the gate-based computers that mark this era of quantum computing \cite{Preskill_2018} have around 100 qubits making this task intractable today. The number of qubits is higher for computers based on quantum annealing, reaching 2000 qubits like the D-Wave computer. However, the correspondence between variables and qubits will not be one to one due to the architecture of these computers, so that we will have a smaller number of useful qubits. The following reference deals with the topology and graph problem mapping on the D-Wave 2000Q QPU computer in detail.

\section{QUBO Model in Quantum Computing}\label{sec:AQC}
Quantum computing as a new computational paradigm can help solve a set of complex problems (routing, scheduling, banking problems, etc.) or tasks that respond with the quantum mechanics' law. However, before solving a problem, we first need to express it in a mathematical formulation that is largely compatible with the underlying physical hardware. This methodology is also useful for quantum computation. One of the framework that allows us to define said mathematical formulation to be solved in a quantum computer is the QUBO.

Adiabatic computation was born from the use of the adiabatic theorem \cite{Born1928, farhi2000quantum} to perform the calculations using the tunnel effect to go from the global minimum of a simple Hamiltonian (A Hamiltonian system is a dynamic system governed by Hamilton equations. In physics, these active systems describe the evolution of a physical system, such as an electron in an electromagnetic field.) \cite{de1956methods,marston1989fourier} to the global minimum of the problem of interest. \\

One of the market leaders for this type of computing is D-Wave, which roughly solves the quadratic unconstrained binary optimization problem (QUBO). The QUBO formulation \eqref{QuboForm} is suitable for running a D-Wave architecture \cite{shin2014quantum}; however, QUBO can be mapped to the Ising \cite{mcgeoch2014adiabatic} model and thus be used in computers based on quantum gates, for example, IBMQ, Rigetti, Xanadu (strawberryfields), etc.

The problems that D-Wave quantum computers are prepared to solve are those that consist in finding the minimum of a function of the following form:

\begin{equation}
    \sum_{i=1}^n b_{i}x_i  + \sum_{i=1}^n \sum_{j=1}^n q_{i,j}x_i x_j,
    \label{QuboForm}
\end{equation}

where the variables $x_{i}\in \{0,1\}$ and the coefficients $b_{i},q_{i,j}\in \mathbb{R}$.

We have then that, given a problem, we need to model it with the above structure where the variables that form the solution will only take the values 0 or 1. Let us observe that, by taking the variables $ x_i $ the values $ 0 $ or $ 1 $, it is true that $ x_i ^ 2 = x_i $. Therefore, we can group the linear terms with the quadratic terms and express the above equation in matrix format:
\begin{equation}
    x^t Q x,
    \label{qbo}
\end{equation}
with $x\in \mathbb \{0,1\}^n$ and $Q \in \mathfrak{M}_{n\times n}$ which is compactly representing the QUBO formulation.
QUBO can be mapped into the Ising model with the change variable: $ z = 1-2x $. Thus, we pass a binary variable $(0,1)$ to a spin variable $(-1,1)$. Thus, given a formulation of a problem to the QUBO, we can implement it and solve it in computers based on quantum gates, only applying the change of variable mentioned.

\section{TSP Formulation} \label{sec:modelTSP}
 As discussed in the introduction, before presenting our GPS model in section four, we will analyze the native TSP model and the MTZ that we aim to improve in terms of the number of variables.

\subsection{Native Formulation} \label{sec:TSPGeneral}
In this section, we will recall the formulation of the native TSP \cite{papalitsas2019qubo}. This modelling, which has been defined in \cite {lucas2014ising}, despite appearing in a very natural way which facilitates its understanding, requires $ N ^ 3 $ variables to be implemented.

The variables that appear in this model are the variables $ x_{i, j, t} $ such that $ i, j  \in \{0, ..., N + 1 \} $ and $ t \in \{0 , ..., N \} $. Let us consider that the variables $ x_ {i, i, t} $ do not exist in this model. \\
The interpretation of the variables $ x_{i, j, t} $ is simple, since $ x_{i, j, t} = 1 $ if at instant $ t $ we traverse the edge that connects the cities $ i $ and $ j $, and $ x_ {i, j, t} = 0 $ for all other cases. \\

We can define the objective function of the native (Native in the sense of general, the most used) TSP\cite{papalitsas2019qubo} as:
 
\begin{equation}
    \sum_{u=0}^{N+1}\sum_{v=0}^{N+1} \sum_{t=0}^N x_{u,v,t} d_{u,v}.
    \label{FunctObjTSPGen}
\end{equation}

where $d_{u,v}$ represents the distance between nodes $u$ and $v$. This objective function is subject to a series of restrictions:

\begin{itemize}
    \item \textit{Constraint 1.} The salesman must leave each city once. 
    \begin{equation}
        \text{For each }u\in \{0,...,N\} \text{: }\sum_{v=1}^{N+1}\sum_{t=0}^{N} x_{u,v,t} = 1.
        \label{res1TSPGen}
    \end{equation}
    
    \item \textit{Constraint 2.} Each city must be reached once. 
    \begin{equation}
        \text{For each }v \in \{1,..,N+1\} \text{: }\sum_{u=0}^N \sum_{t=0}^N x_{u,v,t} = 1.
        \label{res2TSPGen}
    \end{equation}

\item \textit{Constraint 3.} If the salesman leave a city, he cannot return to it later. This constraint ensures that no unconnected cycles are formed as a solution. There are two ways of posing this constraint.
\begin{itemize}
    \item Imposing that once he leaves a city he cannot return to it.\\
    For each $u \in \{1,...,N+1\}$: 
    \begin{equation}
        \sum_{v=0}^{N+1}\sum_{t=0}^{N}\sum_{w=0}^{N+1}\sum_{j=t+1}^N x_{u,v,t}x_{w,u,j} = 0.
        \label{res4TSPGen}
    \end{equation}
    \item Imposing that once he arrives in a city, at the next moment, he must leave it.\\
    For each $t\in \{0,...,N-1\}$, $u,v \in \{0,...,N\}$:
    \begin{equation}
        x_{u,v,t}(1-\sum_{w=1}^{N+1}x_{v,w,t+1}) = 0.
        \label{res5TSPGen}
    \end{equation}
    
\end{itemize}
    
\end{itemize}

This formulation requires $N^3$ variables. Next, we will analyze another model used to define the TSP which is less used in quantum \textit{annealing} articles.

\subsection{MTZ formulation.}  \label{sec:modelMTZ}
Recalling the idea of this formulation is to consider the variables $ x_{i, j} = 1 $ if the edge that connects the cities $ i $ and $ j $ appears in the solution path, where $ x_{i, j} = $ 0 for all other cases. Once we have these variables, we can establish order on the route employing a set of variables that will represent the moment the salesman arrives at that city (the variable $ u_i $ expressed in binary format, will take the integer value $ t $ if the city $ i $ is reached in the $ t^{th}$ position.). This model requires $ N ^ 2log_{2} (N) $, greatly improving the number of variables in the general formulation. However, when implemented using \textit{annealing} it presents surprisingly inaccurate results. This is because the \textit{annealing} algorithm gets stuck trying to minimize the part of the objective function generated by the the sub-tour's constraint \cite{formulacion_MTZ}, since the representation of integers in their binary format has the disadvantage that close numbers such as $ 2 ^ n-1 $ and $ 2 ^ n $ differ by a large number of qubits, so from the \textit{annealing} they are perceived as very different solutions. \\

Once the two most common QUBO models of the TSP have been presented, let us analyze the formulation with which we improve the number of variables of the previous two.

\section{GPS formulation}\label{sec:modelGPS}
Now we are starting to present our work. To develop this model, we take the variables $ x_ {i, j, r} $ with $ i, j \in \{0, ..., N + 1 \} $ and $ r \in \{0,1,2 \} $. In all the modelling, the variables $ x_ {i, j, r} $ such that $ i = j $ are not considered. We work with directional edges, that is, if in the model the edge $(i,j)$ appears, we will understand that first we go through node $i$ and immediately after that we go to $j$. Let us analyze what each variable represents:
\begin{itemize}
    \item  $x_{i,j,0}=1$ means that the edge $ (i, j) $ does not appear in the path and the node $ i $ is reached earlier than the $ j $.
    \item  $x_{i,j,1}=1$ means that the edge $ (i, j) $ appears in the path, so the node $ i $ is reached earlier than the $ j $.
    \item  $x_{i,j,2}=1$ means that the edge $ (i, j) $ does not appear in the path, and the node $ j $ is reached earlier than the $ i $.
\end{itemize}

\begin{figure}[htbp]

	\centering
\includegraphics[width=0.4\textwidth]{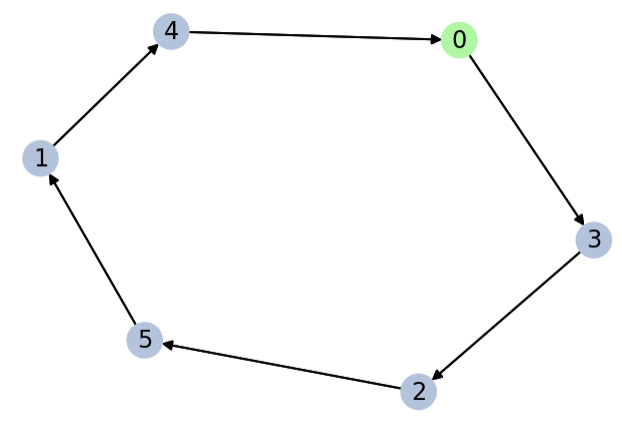}
\caption{Example of a TSP solution with six different cities. It begins at node 0, and the arrows indicate the order in which the towns will be travelled.}
 \label{fig:graph}
\end{figure}

Let us help ourselves with \eqref{fig:graph} to understand this with an example.
previous graph. In this particular case, $ x_{4,5,0} = 0 $ and $ x_{4,5,2} = 0 $ since the edge $ (4,5) $ does appear in the solution. On the other hand $ x_ {4,5,1} $ will also be $0$ because although edge $ (4,5) $ does appear in the graph, node $5$ will be visited before node $4$.

Let us, therefore, see examples in which these variables do not take the value zero:
\begin{itemize}
    \item $ x_{5,1,1} = 1 $: in this case it will take the value $1$ since edge $ (5,1) $ appears in the solution and node $5$ is visited first.
    \item $ x_{4,1,0} = 1 $: this is because in the solution we do have the connection $ (4,1) $ but the node $1$ is visited before node $4$.
    \item $ x_{5,3,2} = 1 $: since the edge $ (3,5) $ does not appear and node $3$ is visited first.
\end{itemize}

From the definition of our variables, we can define the distance travelled through the following objective function as:

    \begin{equation}
    \sum_{i=0}^{N+1} \sum_{j=0}^{N+1} d_{i,j} x_{i,j,1}.  
    \label{FunctObjGPS}
    \end{equation}

The constraints that must be met are:
\begin{itemize}
    \item \textit{Constraint 1}: For each $i,j$ one and only one of the 3 cases of $ r $ must be given, so 
    \begin{equation}
      \label{GPS-Constraint1}
      \text{For all }i,j \text{: }\sum_{r=0}^2 x_{i,j,r} = 1.  
    \end{equation}
    
    \item \textit{Constraint 2}: Each node must be exited once. 
    \begin{equation}
      \label{GPS-Constraint2}
      \text{For each }i \in \{0,...,N\} \text{: }\sum_{j=0}^{N+1} x_{i,j,1} = 1.
    \end{equation}
    
    \item \textit{Constraint 3}: Each node must be reached once.
    \begin{equation}
      \text{For each }j \in\{1,...,N+1\} \text{: }\sum_{i=0}^N x_{i,j,1
    } = 1.  
    \label{GPS-Constraint3}
    \end{equation}
    
    \item  \textit{Constraint 4}: If node $ i $ is reached before $ j $, then node $ j $ is reached after $ i $, so, $ \text{for all}\quad  i, j \in \{0 ,. .., N + 1 \} \quad \text {such that} \quad i \neq j \text {:} $
    
    \begin{equation}
      x_{i,j,2} = 1-x_{j,i,2}.  
      \label{GPS-Constraint4}
    \end{equation}
    It would also have to be specified for $r = 0$ and $r = 1$, however this restriction is sufficient since by \eqref{GPS-Constraint1} it is implicit.
    
    \item \textit {Constraint 5}: If node $ i $ is reached before node $ j $ and node $ j $ is reached before node $ k $, then node $ i $ must be reached before $ k $. This condition will prevent the route from returning to a node from which it had already exited, thus preventing cycles from forming. We then arrive at the penalty function equation \eqref{GPS-Constraint5}. \\
    \begin{equation}
    \begin{aligned}
      &\sum_{i=1}^N\sum_{j=1}^N\sum_{k=1}^N  (x_{j,i,2}x_{k,j,2} - x_{j,i,2}x_{k,i,2} \\
      & -x_{k,j,2}x_{k,i,2} +x_{k,i,2}).
      \label{GPS-Constraint5}
      \end{aligned}
    \end{equation}
    With only the cases in which $ i \neq j $, $ i \neq k $ and $ j \neq k $ will be taken in the summation and in the annex \eqref{sec:GPSpenalization} we will provide the approach followed to arrive at it. \\
The following is deduced from the equation \eqref{GPS-Constraint5}. We have $ x_ {i, j, 2} = 0 $ if $ i $ is reached before $ j $ and $ x_ {i, j, 2} = 1 $ in the case where $ j $ is reached before $ i $. 
Thus, with the previous equation we penalize these following cases in which $ x_ {i, j, 2} = 0, x_ {j, k, 2} = 0 $ and $ x_ {i, k, 2} = 1 $ and $ x_ {i, j, 2} = 1, x_ {j, k, 2} = 1 $ and $ x_ {i, k, 2} = 0 $ which lead to cases in which it would be forming cycles (for these two situations we have that the value of the parentheses is $ 1 $ and for the rest $ 0 $). In \eqref{QTSP-Constraint10} we have a similar situation for our VRP formulation, where we offer more details.
For this condition, we must have directly constructed a penalty function that avoids erroneous cases without first posing linear conditions through which to generate its corresponding penalty function.

\end{itemize}

Formulated in this way we have managed to reduce the number of variables required from $N^2 \log_{2}{N}$ to $3N^2$, achieving very noticeable reductions when working with large problems. Once we have this formulation, let us see how we can generalize it to the new VRP formulation.

\section{New VRP Formulation}\label{sec:modelVRP}
This section develops our VRP into the QUBO model using the GPS formulation.

As discussed in the introduction, this model is optimal concerning the number of binary variables used. However, this generalization does not appear as naturally as expected because it requires a delicate step to get the constraints of the equation \eqref{QTSP-Constraint10}. To do this, we will detail each step and explain each constraint step by step.

\subsection{Original formulation $5N^2Q$}\label{sec:VRPGeneral}
For this new VRP, we will consider that $ N $ is the number of cities and $ Q $ is the number of available vehicles. We first present the variables that will form the problem. We then take the following set of variables. 
\begin{equation}
    \label{variablesVRPGEN}
    \begin{aligned}
& x_{i,j,r,q} \text{ with }i,j \in \{0,...,N+1\}, \\
& r\in \{0,1,2,3,4\} \text{ and }q \in \{1,...,Q\}
\end{aligned}
\end{equation}
    
In all the modelling, the variables $ x_ {i, j, r} $ such that $ i = j $ are not considered. The variables $ i $, $ j $ refer to the cities must travel to, and the variable $ q $ refers to the vehicle.
The nodes $ 0 $ and $ N + 1 $ correspond to the starting and ending points. Note that they may be the same node but we will separate them for convenience in the formulation.
The values $ d_ {i, j} $ with $ i, j \in \{0, ..., N + 1 \} $ correspond to the distance between node $ i $ and $ j $.
Let us dive into the interpretation of each variable:

\begin{itemize}
    
    \item  $x_{i,j,0,q}=1$ means that the vehicle $ q $ travels to the cities $ i $ and $ j $, does not travel across the edge $ (i, j) $ and arrives at the city $ i $ before the $ j $.
    
    \item  $x_{i,j,1,q}=1$ means that the vehicle $ q $ travels to the cities $ i $ and $ j $ travels across the edge $ (i, j) $ (that is, once it passes through the city $ i $ the next city it reaches is the $ j $) and therefore the city $ i $ is reached earlier than the city $ j $.
    
    \item  $x_{i,j,2,q}=1$ means that the vehicle $ q $ travels through the cities $ i $ and $ j $ and arrives at the city $ j $ earlier than at the $ i $.
    
    \item  $x_{i,j,3,q}=1$ means that the vehicle $ q $ does not go through the cities $ i $ and $ j $, and the city $ i $ is reached earlier than the city $ j $. Note that $ x_ {i, j, 3, q} $ can take the value $ 1 $ whether the vehicle $ q $ passes through one of both cities or neither of them.

    \item  $x_{i,j,4,q}=1$ means that the vehicle $ q $ does not travel to the cities $ i $ and $ j $, and the city $ j $ is reached earlier than the city $ i $.

\end{itemize}
Even if no vehicle passes through the objects $ i $ and $ j $, the formulation must establish an order between them. However, this restriction does not make the modelling meaningless, since we can assume that if the vehicles are ordered in the order of $ \{1, ..., Q \} $, then $ i $ will be reached before $ j $ if the vehicle that passes through node $ i $ has a lower number than the one that passes through node $ j $. 
Once the interpretation of each variable is explained, let us analyse the constraints that must be met.
\begin{itemize}
    \item \textit{Constraint 1:} For each $i,j,q$, one and only one of the possibilities must be met for $ r $, so:
    \begin{equation}
      \text{For all }i,j,q \text{: } \sum_{r=0}^4 x_{i,j,r,q}=1, 
      \label{VRP:constrain1}
    \end{equation}
    
    \item \textit{Constraint 2:} Each vehicle has to fulfill that it leaves the starting position. For this situation, we are going to impose that:
    \begin{equation}
      \text{For all }q \text{: } \sum_{j=1}^{N+1} x_{0,j,1,q}=1, 
      \label{VRP:constrain2}
    \end{equation}
   
   No vehicle can return to the starting position from a city, so: 
    \begin{equation}
      \text{For all }q \text{: } \sum_{i=0}^{N+1}x_{i,0,1,q} = 0,  
    \end{equation}
    
    \item \textit{Constraint 3:} Every vehicle must finish in the final position. For this, it must be fulfilled that:
    \begin{equation}
      \text{For all }q \text{: } \sum_{i=0}^N x_{i,N+1,1,q}=1,
      \label{VRP:constrain3}
    \end{equation}

    No vehicle can leave the final position. We then have that:
    \begin{equation}
      \text{For all }q \text{: } \sum_{j=0}^{N+1}x_{N+1,j,1,q} = 0.  
    \end{equation}
   
    \textbf{Vehicles that do not travel on any road} will meet all constraints when taking the following condition:  
    $$x_{0,N+1,1,q}=1.$$
   
    \item \textit{Constraint 4:} The vehicle must leave once and only once from each city, then:
    \begin{equation}
      \text{For each }i \in \{1,...,N\} \text{: }\sum_{q=1}^Q\sum_{j=1}^{N+1} x_{i,j,1,q} = 1.
      \label{VRP:constrain4}
    \end{equation}

    \item \textit{Constraint 5:} The vehicle must arrive once and only once to each city, then:
    \begin{equation}
      \text{For each }j \in \{1,...,N\}\text{: }\sum_{q=1}^Q \sum_{i=0}^N x_{i,j,1,q} = 1.  
      \label{VRP:constrain5}
    \end{equation}
    
    \item \textit{Constraint 6:}  The city $ i $ is reached before the city $ j $ does not depend on each vehicle. Therefore, for all the vehicles that either arrive at city $ i $ earlier than $ j $, or arrive at city $ j $ earlier than $ i $. Introducing the auxiliary variables $ a_ {i, j} $, we have the following constraint. $ \text {For all} \quad i, j \in \{1, ..., N \} \text {:} $
    \begin{equation}
      \sum_{q=1}^Q x_{i,j,0,q}+x_{i,j,1,q}+x_{i,j,3,q} = a_{i,j} Q.  
      \label{rest7}
    \end{equation}
    
    It will then be true that for each $ i, j $ or $ a_ {i, j} = 1 $, which means that the city $ i $ is reached earlier than the city $ j $ and therefore for each $ q $ we will have $ x_ {i, j, r, q} = 1 $ for any value of the $ r $  in which $ i $ is reached before $ j $, or $ a_ {i, j} = 0 $, and, we will have $ x_{i, j, r, q} = 0 $ for all the vehicles and for values $ r $ where $ i $ is reached before $ j $.
    

    \item \textit{Constraint 7:} If the vehicle $ q $ arrives in the city $ j $, then the vehicle $ q $ must leave the city $ j $. For this we impose the constraint that $ \text {for} \quad i \in \{0, ..., N \} \quad \text {,} \quad j \in \{1, ..., N \} \quad \text{and } \quad q \in \{1, ..., Q \} \text {:} $
    \begin{equation}
      \label{QTSP-Constraint888}
      x_{i,j,1,q}(1-\sum_{k=1}^{N+1} x_{j,k,1,q})= 0. 
    \end{equation}
    
\end{itemize}

Let us now impose the conditions that make vehicles run on a tour.
\begin{itemize}
    
    \item \textit{Constraint 8:} It must be fulfilled that either the vehicle pass through the city $ i $ before the $ j $ or arrive before to the city $ j $ than the $ i $. Therefore, it must be verified that, $ \text {for} \quad i \in \{0, ..., N \} \text {,} \quad j \in \{1, ..., N \} \quad \text {and} \quad q \in \{1, ..., Q \} $:
    \begin{equation}
    \begin{aligned}
        & x_{i,j,0,q}+x_{i,j,1,q}+x_{i,j,3,q}=1-(x_{j,i,0,q}\\
        &+x_{j,i,1,q}+x_{j,i,3,q}).
    \end{aligned}
    \end{equation}
    
    \item \textit{Constraint 9:} If city $ i $ is reached before $ j $ and city $ j $ is reached before city $ k $, then city $ i $ must be reached before city $ k $. This condition will prevent the vehicle from returning to a city it has already passed through and therefore prevents a cycle from forming. \\
     To introduce this constraint, we will directly calculate a penalty function worth $ 0 $ in the correct cases and $ 1 $ in those that are not. To facilitate the understanding of the penalty function, we are going to take, for $ i, j, k, q $, the following variables: \\
    \begin{equation}
    \label{variables}
    \begin{aligned}
      & a_{i,j}=x_{i,j,0,1}+x_{i,j,1,1}+x_{i,j,3,1} \\  & a_{j,k}=x_{j,k,0,1}+x_{j,k,1,1}+x_{j,k,3,1} \\ & a_{i,k}=x_{i,k,0,1}+x_{i,k,1,1}+x_{i,k,3,1}.
    \end{aligned}
    \end{equation}
    Remember that it is not necessary to introduce these conditions because the constraint \eqref{rest7} establishes the correct values of the variables $a_{i,j}$.
    Therefore, $ a_ {i, j} = 1 $ means that the city $ i $ is reached before the city $ j $ and the same with $ j $ and $ k $. Also, it is very important to remember that due to the same constraint \eqref{rest7}, we can take any of the vehicles as a reference. In this case, we have taken the first vehicle as a reference. \\
   
In this way, fixed $ i, j, k $, we have the 3 variables $ a_{i,j}, a_{j,k}, a_{i,k} $. Remember that $ a_{i,j}, a_{j,k}, a_{i,k} $ only take the values $ 0 $ or $ 1 $. Also, let us note that the cases that lead to values of the variables for which cycles can be formed and that we must discard are $ (a_{i,j}, a_{j,k}, a_{i,k}) = (0,0,1) $ and $ (a_{i,j}, a_{j,k}, a_{i,k}) = (1, 1,0) $. \\
    
In the case $ (0,0,1) $ we would have that the city $ j $ is reached after the $ i $, the $ k $ after the $ j $, and yet the city $ k $ is reached rather than $ i $, which is absurd. The case $ (1,1,0) $ cannot be given either, since it reaches $ i $ before $ j $ and $ j $ before $ k $, so it cannot be that we also reach $ k $ before $ i $. We therefore must construct a penalty function so that for $f (a_{i,j}, a_{j,k}, a_{i,k})  $ it holds that $ f (0,0,1)> 0 $, $ f (1,1,0)> 0 $ y $ f (a_{i,j}, a_{j,k}, a_{i,k}) = 0 $ for all other cases. A function that satisfies these conditions is from the equation \eqref{QTSP-Constraint1000}.

     \begin{equation}
      \label{QTSP-Constraint1000}
      \begin{aligned}
        & f(a_{i,j},a_{j,k},a_{i,k}):= a_{i,j}a_{j,k} - a_{i,j}a_{i,k} \\
        & -a_{j,k}a_{i,k} +a_{i,k}^2.
      \end{aligned}
    \end{equation}
    
   then, adding to the cost function the equation \eqref{QTSP-Constraint10}
    \begin{equation}
      \label{QTSP-Constraint10}
      \begin{aligned}
      &\lambda \sum_{i=1}^N\sum_{j=1}^N\sum_{k=1}^N (a_{i,j}a_{j,k} \\
      & - a_{i,j}a_{i,k} -a_{j,k}a_{i,k} +a_{i,k}^2),  
      \end{aligned}
    \end{equation}
    we will have that the best solutions will be those that comply with this constraint.
    
    \item \textit{Constraint 10:} The objective we seek is to minimize vehicle travel time. What we could do is to see how long each vehicle takes to complete the route and try to minimize as much of the time as possible. However, this function soon becomes complex so we have decided to develop a different idea that simplifies the process and smoothes the objective function. If we impose the condition that all vehicles travel less distance than the distance travelled by vehicle number $1$, we will have that minimizing the maximum of the distances will be equivalent to minimizing the distance travelled by the first vehicle. We then have the following condition. For each $q\in \{2,...,Q\}$:
    \begin{equation}
       \sum_{i=0}^{N+1} \sum_{j=0}^{N+1} d_{i,j}x_{i,j,1,q} \leq \sum_{i=0}^{N+1} \sum_{j=0}^{N+1} d_{i,j}x_{i,j,1,1}
       \label{VRP:constrain11}.  
    \end{equation}
    We transform this inequality into equality by taking once again $D_{max}:=\sum_{i=0}^n \max_j \{d_{i,j}\}$ and the variables $b_{h,q}$ (the variables $b_{h,q}$ are like the sub tour's one in the MTZ slack variables and they are in their binary expression) in:
    \begin{equation}
    \begin{aligned}
    & \sum_{i=0}^{N+1}\sum_{j=0}^{N+1} d_{i,j}x_{i,j,1,q} +\sum_{h=0}^{h_{max}}2^h b_{h,q}\\
    & -\sum_{i=0}^{N+1} \sum_{j=0}^{N+1} d_{i,j}x_{i,j,1,1} = 0.  
    \end{aligned}
    \end{equation}
    Under these conditions the function to be minimized corresponds to:
    \begin{equation}
        \sum_{i=0}^{N+1} \sum_{j=0}^{N+1} d_{i,j}x_{i,j,1,1}.
    \end{equation}
    This condition has the disadvantage that we are eliminating solutions where it is another vehicle that travels the longest distance. Let us explore how to avoid this problem and get more flexibility in the model to make it easier for the Quantum Annealing to find the optimum one. We can establish an auxiliary variable $D$ and we set that the distance travelled by each vehicle must be less than this variable, that is to say:
    \begin{equation}
    \sum_{i=0}^{N+1} \sum_{j=0}^{N+1} d_{i,j}x_{i,j,1,q} \le D
        \text{ , for all }q \in \{1,...,Q\}.
        \label{QTSP-Constraint11-S}
    \end{equation}
    The variable $D$ is an integer, so we must treat it in some way in order to include it in the model. As we explained in the introduction of the section dedicated to the formulation of the MTZ model (\ref{sec:modelMTZ}), it is convenient to try to avoid the binary representation of integer variables. To do so, we can express $D$ as a combination of the distances between edges by taking $D = \sum_{i=0}^{N+1}\sum_{j=0}^{N+1} x_{i,j}b_{i,j}$. Thus after imposing the constraint \eqref{QTSP-Constraint11-S} we have that the function to minimize is $D$.
    
\end{itemize}
Thanks to this modelling of the new VRP we have been able to reduce the number of variables required to the order of $5N^2Q$. However, we have managed to reduce it even further to $3N^2Q$, which is detailed in  \eqref{sec:modelVRPenhanced}. However, we have preferred to present this other model due to its easy understanding.

\section{Results} \label{sec:resultados}

To test the correct VRP model developed in QUBO, which minimizes the maximum distance that all the vehicles travel, we will present some comparisons of the results obtained through the simulator of the different models that have been discussed in this paper.\\

The code has been implemented on the Ocean library \cite{teplukhin2020electronic} from D-Wave in python. The reader can find the code at \cite{GPSCODE}. 

\begin{figure*}[t!]
\centering
\includegraphics[height=4cm]{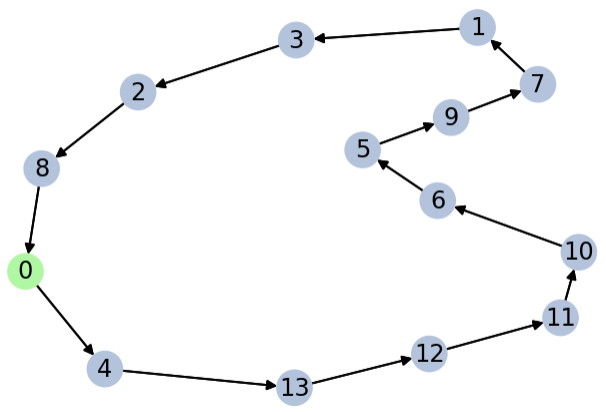}
\includegraphics[height=4cm]{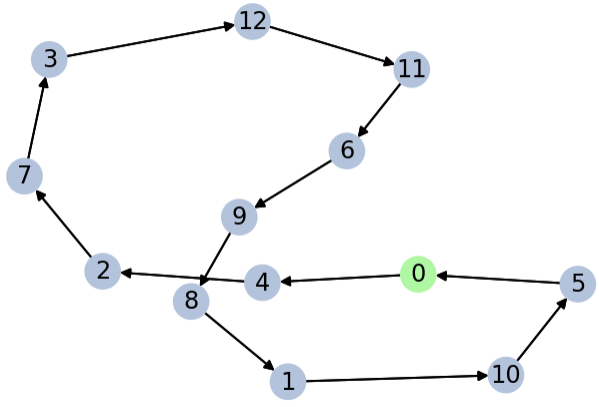}
\includegraphics[height=4cm]{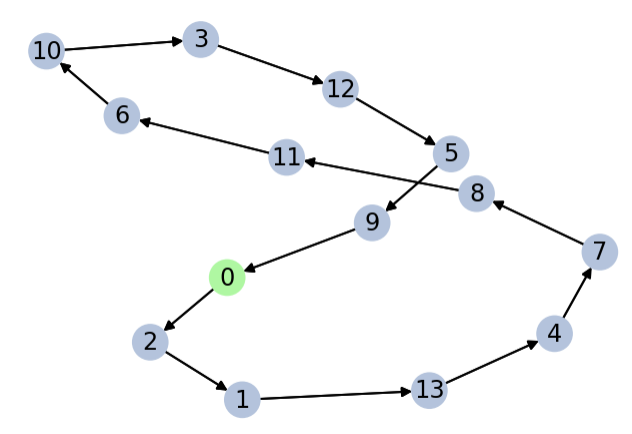}
\includegraphics[height=4cm]{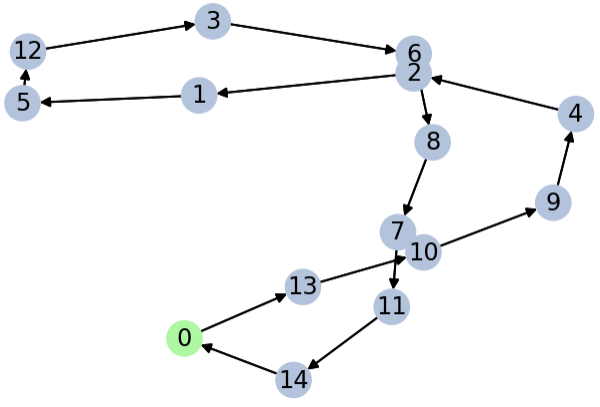}
\includegraphics[height=4cm]{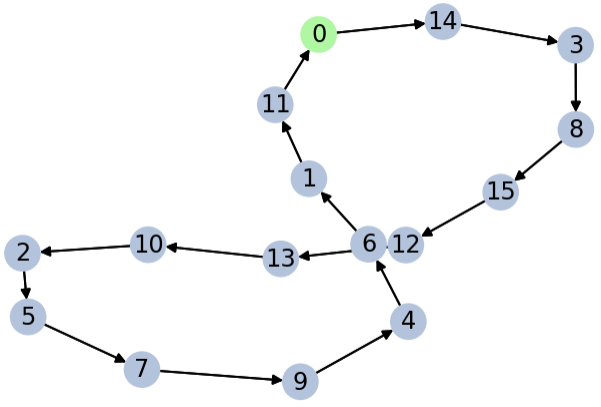}
\includegraphics[height=4cm]{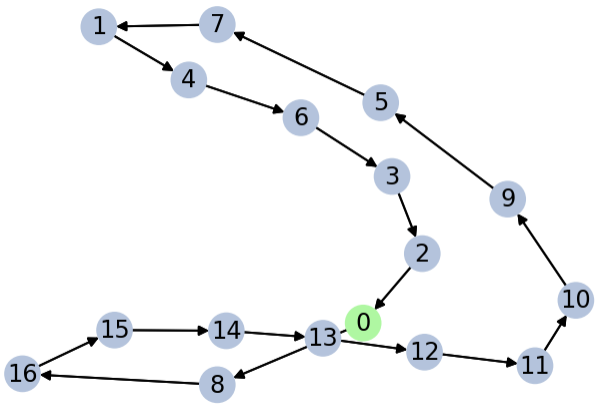}
\caption{In these graphs, we can observe the algorithm's results in different scenarios of the GPS formulation. We can follow the correct scalability of the algorithm. We provide the code \cite{GPSCODE} to check its proper functioning and if someone wants to simulate lower values or values higher than N = 16.}
\label{fig:Results_GPS}
\end{figure*}

Figure \eqref {fig:Results_GPS} offers a sample of our GPS formulation's results when using the D-Wave solver in different scenarios. We highlight some important cases that help us see the good functioning of the algorithm. 

\begin{figure*}[t!]
\centering
\includegraphics[height=4cm]{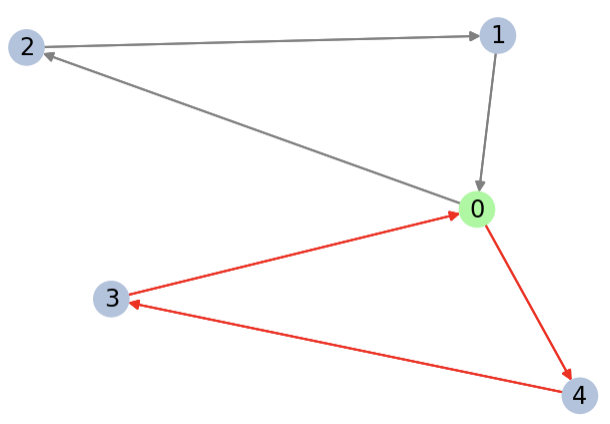}
\includegraphics[height=4cm]{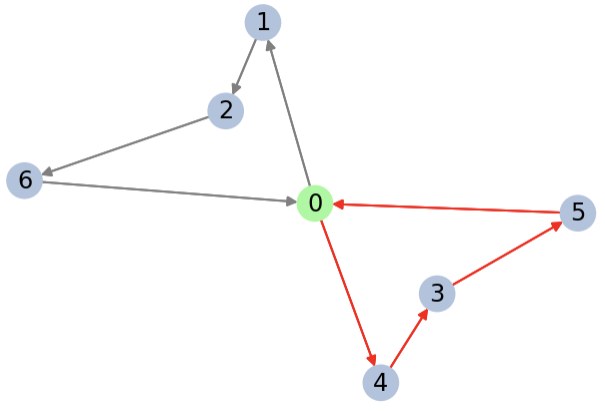}
\includegraphics[height=4cm]{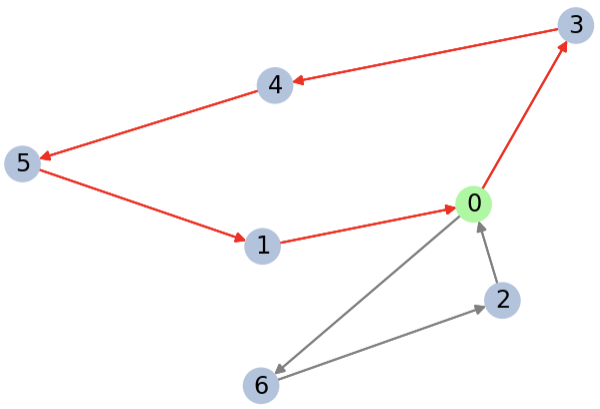}
\includegraphics[height=4cm]{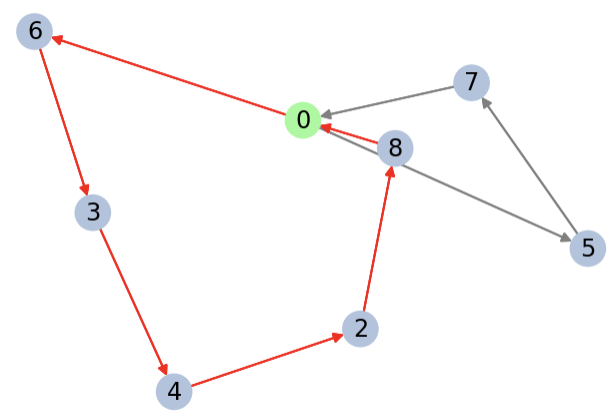}
\includegraphics[height=4cm]{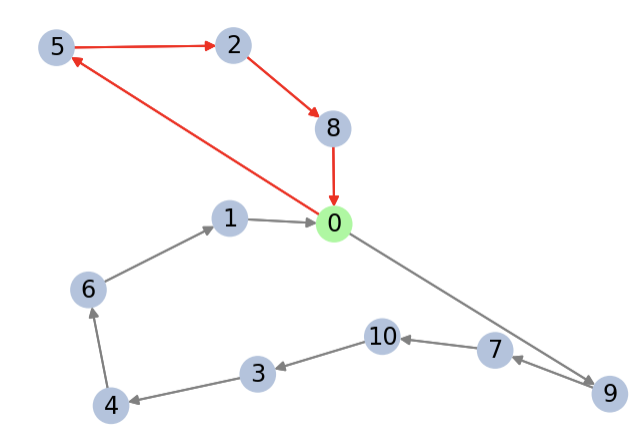}
\includegraphics[height=4cm]{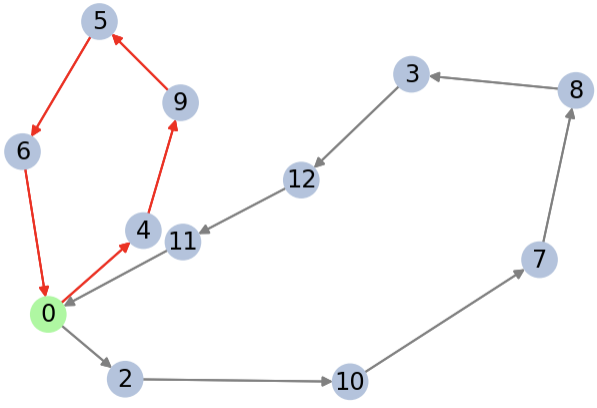}
\caption{In these graphs, we can observe the algorithm's results in different scenarios of the VRP formulation. We can follow the correct scalability of the algorithm. We provide the code \cite{GPSCODE} to verify the proper functioning of the formulation. Vehicle number $ 1 $ is red, and the next is light-steel-blue. While the depot is the $ 0 $ node in pale-green colour, and the rest are represented in light-steel-blue.
In this case, we have variables cities from $ 4 $ to $12$ and using up to $ 2 $ vehicles.
It is important to highlight that this VRP minimises the time travelled by the cars. The number of qubits used is $2418$ to test the last case.}
\label{fig:Results_VRP_GPS}
\end{figure*}

Figure \eqref {fig:Results_VRP_GPS} offers a sample of our VRP formulation's results based on the GPS when using the D-Wave solver in different scenarios. We highlight some important cases that help us see the good functioning of the algorithm. It is important to note that our algorithm minimizes the maximum distance travelled by all the vehicles (this is equivalent to reducing the time spanned by all cars). It is worth mentioning that the number of the qubits needed in the case $N = 8$ and $Q = 3$  is $1778$. Where $N$ is the number of cities and $Q$, the vehicles. In the discussion section, we will analyze this point and its impact on the topology of the QPU architecture and in this case of the D-Wave.

Let us observe in the different tables the comparison of the number of qubits, time during which the D-Wave Quantum Annealing simulator has been executed, and the length of the path found. The sign \text{"-"} represents that the algorithm did not find a possible way during the elapsed time (in minutes). In this examples, the cities which form the TSP to solve are the vertex of the regular polygon with these number of vertex.

\begin{table}[th!]
    \centering
    \begin{tabular}{c|c|c|c}
         \textbf{}   &  \textbf{GPS}   & \textbf{Native TSP} & \textbf{MTZ} \\
         \hline
         Number of qubits & 75 & 100 & 140  \\
         Elapsed Time (min) &  0.332 & 0.08 & 0.569 \\
         Path Length (m) &  5.65 &  5.65 &  5.65 \\
    \end{tabular}
    \caption{A regular polygon layout has been taken where the cities occupy the positions of the nodes \cite{GPSCODE} for the elaboration of all tables. In this scenario of 4 cities, we set comparison with the 3 models, MTZ, native TSP and GPS. The comparison is based on the number of times to find the solution, the distance travelled, and the number of qubits. We can appreciate the good performance of our GPS model, and above all the savings it offers us in the number of qubits.}
    \label{tab:4cuidades}
\end{table}

\begin{table}[th!]
    \centering
    \begin{tabular}{c|c|c|c}
         \textbf{}   &  \textbf{GPS}   & \textbf{Native TSP} & \textbf{MTZ} \\
         \hline
         Number of qubits & 147 & 294 & 266  \\
         Elapsed Time (min) &  0.337 & 0.39 & 1.338 \\
         Path Length (m) &  6.00 &  6.00 &  8.46 \\
    \end{tabular}
    \caption{A regular polygon layout has been taken where the cities occupy the positions of the nodes \cite{GPSCODE} for the elaboration of all tables. In this scenario of 6 cities, we set comparison with the 3 models, MTZ, native TSP and GPS. The comparison is based on the number of times to find the solution, the distance travelled, and the number of qubits. We can appreciate the good performance of our GPS model, and above all the savings it offers us in the number of qubits.}
    \label{tab:6cuidades}
\end{table}

\begin{table}[th!]
    \centering
    \begin{tabular}{c|c|c|c}
         \textbf{}   &  \textbf{GPS}   & \textbf{Native TSP} & \textbf{MTZ} \\
         \hline
         Number of qubits & 243 & 648 & 522  \\
         Elapsed Time (min) &  1.209 & 1.177 & 2.676 \\
         Path Length (m) &  6.122 &  9.58 &  11.46 \\
    \end{tabular}
    \caption{
A regular polygon layout has been taken where the cities occupy the positions of the nodes \cite{GPSCODE} for the elaboration of all tables. In this scenario of 8 cities, we set comparison with the 3 models, MTZ, native TSP and GPS. The comparison is based on the number of times to find the solution, the distance travelled, and the number of qubits. We can appreciate the good performance of our GPS model, and above all the savings it offers us in the number of qubits.}
    \label{tab:8cuidades}
\end{table}

\begin{table}[th!]
    \centering
    \begin{tabular}{c|c|c|c}
         \textbf{}   &  \textbf{GPS}   & \textbf{Native TSP} & \textbf{MTZ} \\
         \hline
         Number of qubits & 363 & 1210 & 770  \\
         Elapsed Time (min) &  3.316 & 3.087 & 4.175 \\
         Path Length (m) &  12.51 &  10.978 &  \text{-} \\
    \end{tabular}
    \caption{A regular polygon layout has been taken where the cities occupy the positions of the nodes \cite{GPSCODE} for the elaboration of all tables. In this scenario of 10 cities, we set comparison with the 3 models, MTZ, native TSP and GPS. The comparison is based on the number of times to find the solution, the distance travelled, and the number of qubits. We can appreciate the good performance of our GPS model, and above all the savings it offers us in the number of qubits.}
    \label{tab:10cuidades}
\end{table}

\begin{table}[th!]
    \centering
    \begin{tabular}{c|c|c|c}
         \textbf{}   &  \textbf{GPS}   & \textbf{Native TSP} & \textbf{MTZ} \\
         \hline
         Number of qubits & 507 & 2028 & 1066  \\
         Elapsed Time (min) &  7.992 & 9.677 & 10.578 \\
         Path Length (m) &  14.286 &  12.28 &  \text{-} \\
    \end{tabular}
    \caption{A regular polygon layout has been taken where the cities occupy the positions of the nodes \cite{GPSCODE} for the elaboration of all tables. In this scenario of 12 cities, we set comparison with the 3 models, MTZ, native TSP and GPS. The comparison is based on the number of times to find the solution, the distance travelled, and the number of qubits.}
    \label{tab:12cuidades}
\end{table}

These results have been obtained using a simulator because we would require access to a quantum computer for a time similar to that needed to perform the simulations (in some cases more than an hour). However, it is the benefits of modelling with few qubits (such as GPS modelling) will be much more notable when these problems are implemented in real quantum computers. Other studies that did not require many hours of the quantum computer were carried out on the D-Wave\_2000Q\_6. In the discussion section, we detail some interesting cases.

\subsection{Discussions}\label{sec:discussions}
\begin{figure}[htbp]
	\centering
		\includegraphics[width=0.4\textwidth]{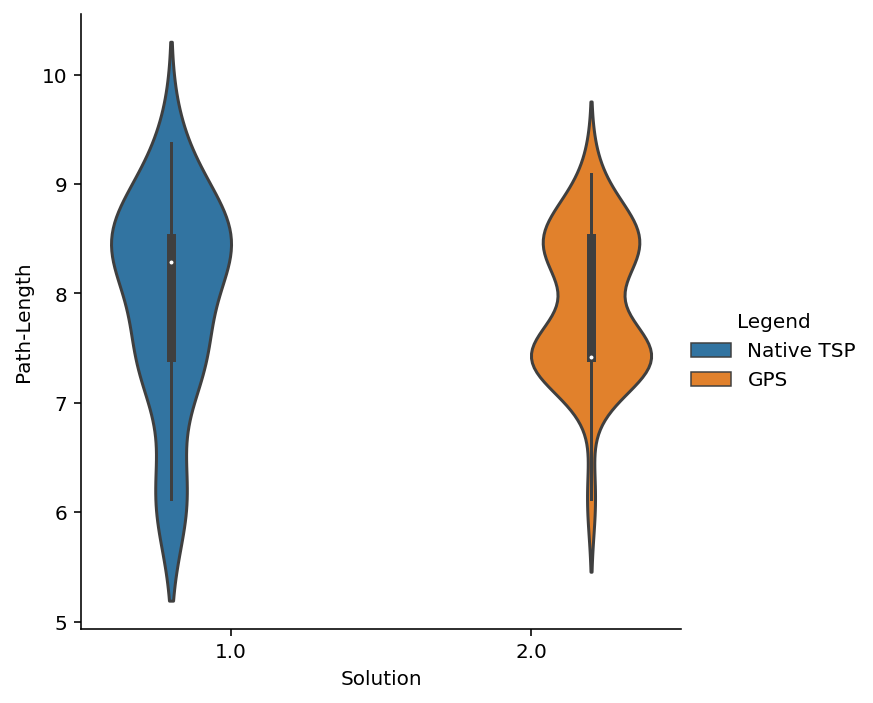}
		\caption{Path length comparison for N = 9. In this graph, we see how the length of the solution paths for the case of 9 Cities is very similar so that both models give good results.}
		\label{fig:Capcodigo3Long}
\end{figure}

\begin{figure}[htbp]
	\centering
		\includegraphics[width=0.4\textwidth]{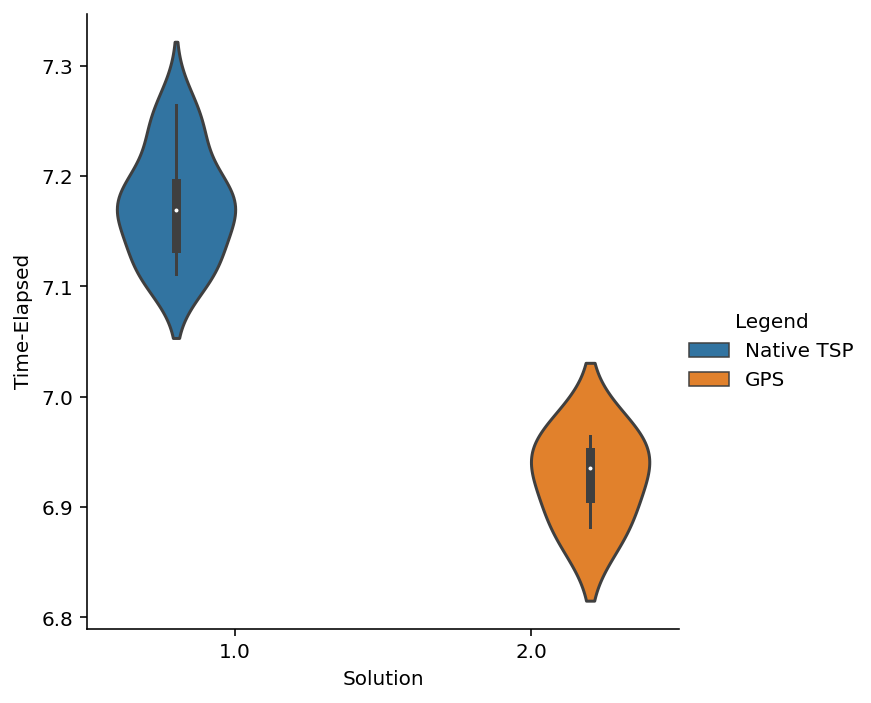}
	\caption{Time comparison for N = 11. For the example of 11 cities, we can observe that the outcomes are quite similar because although there is a mean difference of about 20 seconds between the results of both simulations, the experience with this problem and other similar ones is that this very small difference does not affect the results on the length of the solution path. }
	 \label{fig:Captura3t11}
\end{figure}

Once the different models had been implemented, we achieved the following results.
Through the results of the figure \eqref{fig:Capcodigo3Long} up to the figure \eqref{fig:Captura3t11}, the good performance of our formulation compared to the general TSP \cite{papalitsas2019qubo} can be observed. An almost identical operation is seen with the generic TSP, except that we are improving at least the number of qubits for the same cases in our proposal. Although the time difference is not significant again, the difference between path lengths is. Let us remember that the advantage of the formulation in which we have worked is based on improving the number of qubits used. We then have that the larger the problems we are working on, the better this difference will be appreciated in the number of variables.

The MTZ model does not offer positive results. This is since \textit{Annealing} presents many difficulties to find minimum expressions in which the representation of integers appears in their binary format. This is because although the numbers $ 2 ^ k-1 $ and $ 2 ^ k $ are close, they are not close in their binary form since they differ in $ k $ variables, so the \textit{annealing} tends to present bad results. Apart from that adjusting, the Lagrange coefficient of this type of constraint is also a complicated task. \\
    
Native TSP and GPS modelling show better results. While it is true that general modelling gives slightly better results, it requires the use of a higher number of qubits. This may be since the function to be optimized for this model has a smaller number of local minima where the \textit {Annealing} can get stuck or to a bad fit of the Lagrange coefficients.

The problem on which the simulations are carried out consists in finding the optimal path when the points are placed on the vertices of the regular polygons that have the same number of vertices as nodes in our problem. \\


\begin{figure}[htbp]
	\centering
		\includegraphics[width=0.4\textwidth]{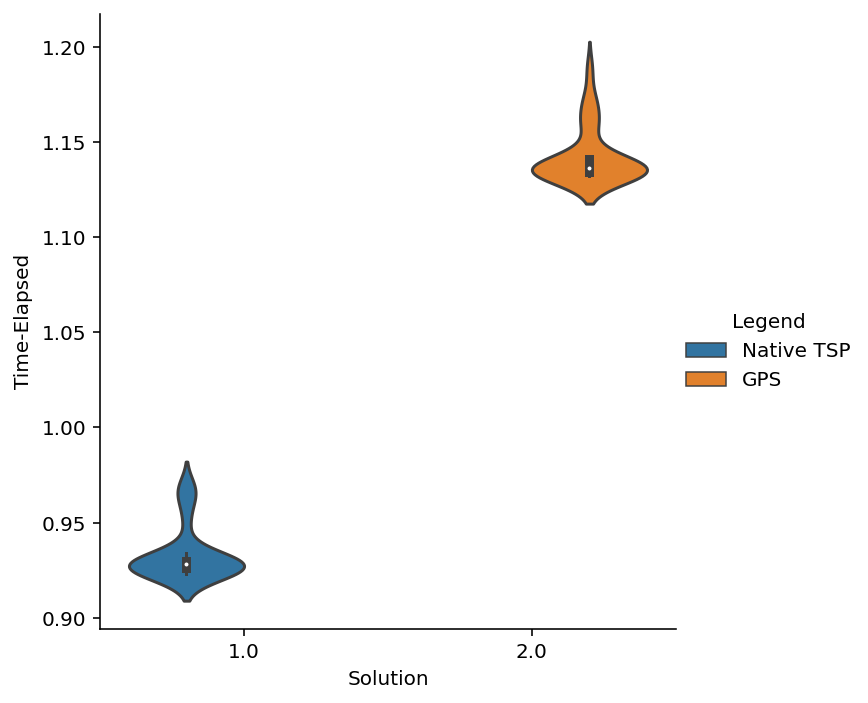}
		\caption{Time comparison for N = 9. This graph shows the time taken to carry out the executions in the case of 9 cities. Although it seems that there is a lot of difference, it only represents 10\% of the total time, which, as we have seen in other experiences, is not significant.}
		\label{fig:Captura_codigo3}
\end{figure}


\begin{figure}[htbp]
	\centering
		\includegraphics[width=0.4\textwidth]{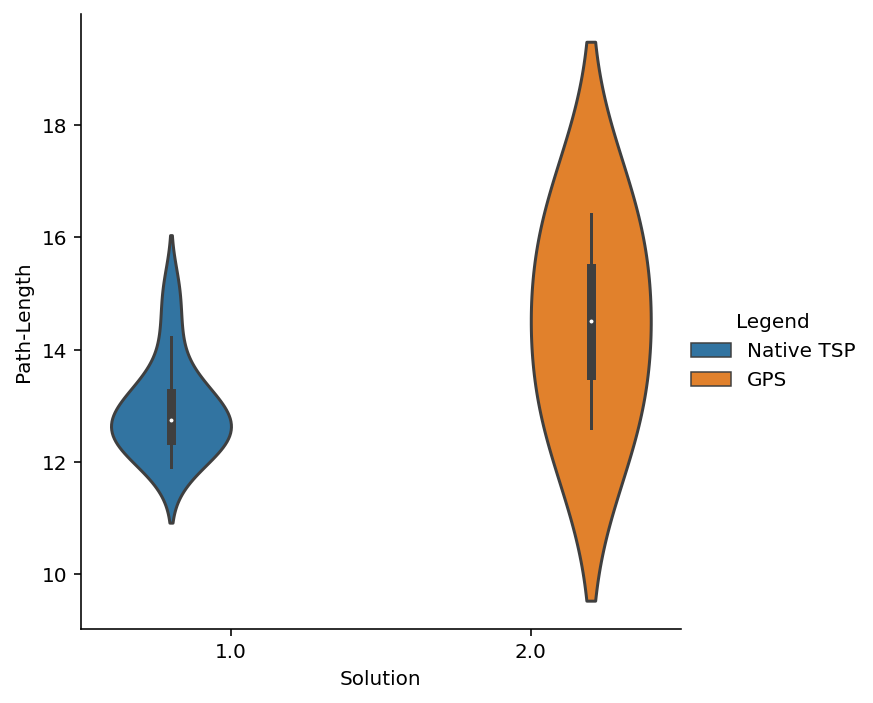}
		\caption{Path length comparison for N = 11. For the example of 11 cities, we can observe that the outcomes are quite similar. Although the time difference is not significant again, the difference between path lengths is. Let us remember that the advantage of the modelling we have worked is based on improving the number of qubits used. We then have that the larger the problems we are working on, the better that difference will be appreciated in the number of variables.}
		\label{fig:Captura_codigo3long11}
\end{figure}

\begin{figure}[h!]
	\centering
	\includegraphics[width=0.4\textwidth]{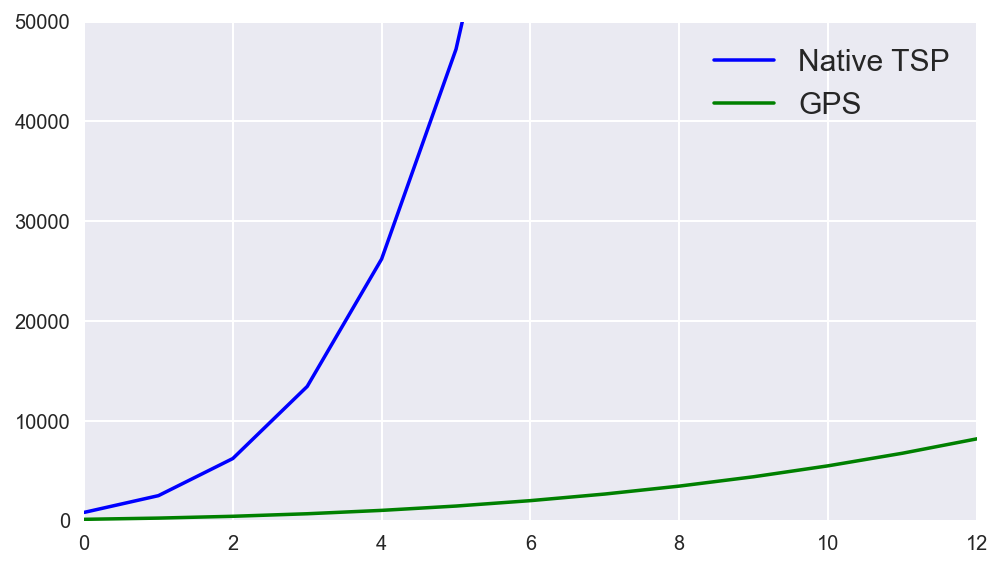}
	\caption{In this figure we can appreciate the exponential behaviour and the number of interconnections that each model offers. Our model (GPS) improves the number of qubits and gives us a great result reducing the number of connections a lot. The Native\_TSP behaves as $0.8 (N + 2) ^ 5$ while the GPS as $ 2 (N + 2) ^ 3 $. }
	\label{fig:benchVertex}
\end{figure}

One of the behaviours and results that we believe is important to mention is the following. We realized that it is even more important to consider the number of edges that our model generates. The vertex/connections in a quantum computer are limited and define our quantum computer's typology and quality for error mitigation. Thus, a model that produces many edges (direct links) may request more from a computer than another generates few. The figure \eqref{fig:benchVertex} offers us a comparative study between our GPS model and the native TSP. This figure shows the exponential behaviour and the number of interconnections that each model offers. Our model improves the number of qubits and gives us a great result reducing the number of connections a lot. The native TSP behaves as $0.8 (N + 2) ^ 5$ while the GPS as $ 2 (N + 2) ^ 3 $.

One aspect of GPS worth commenting on here is to generalize it also to be used for the Cutting-plane method. We must change the current constraint \eqref{GPS-Constraint5} since this methodology only works with linear constraints.
The way to do this is as follows.
For each $i,j,k$:
\begin{itemize}
    \item  $x_{j,i,2} + x_{k,j,2} \leq 2x_{k,i,2} + w_{i,j,k}^1$ 
    \item  $x_{j,i,2} + x_{k,j,2} \geq 2x_{k,i,2} - w_{i,j,k}^2$.
\end{itemize}
In these equations, the variables $w_{i,j,k}^p$ are auxiliaries. The purpose of these variables is to satisfy the said constrains. These two restrictions are satisfied by all cases of $(x_{j, i, 2}, x_{k, j, 2}, x_{k, i, 2})$ except for $(0, 0, 1)$ (because it  doesn't satisfy the second constraint) and $(1, 1, 0)$ (because it doesn't satisfy the first constraint).

\section{Conclusions and Further Work}\label{sec:conclusions}

The importance of finding a good formulation in the QUBO model that minimizes the number of variables to be used is crucial for the computing era we are in, as we have commented throughout this work. It is true that, although the technology of annealing-based quantum computers allows having much more qubits than gate-based computers, it remains a limitation and, therefore, a challenge to try to solve. Hence highlighting the importance of our research.

\begin{figure}[ht!]
	\centering
	\includegraphics[width=0.4\textwidth]{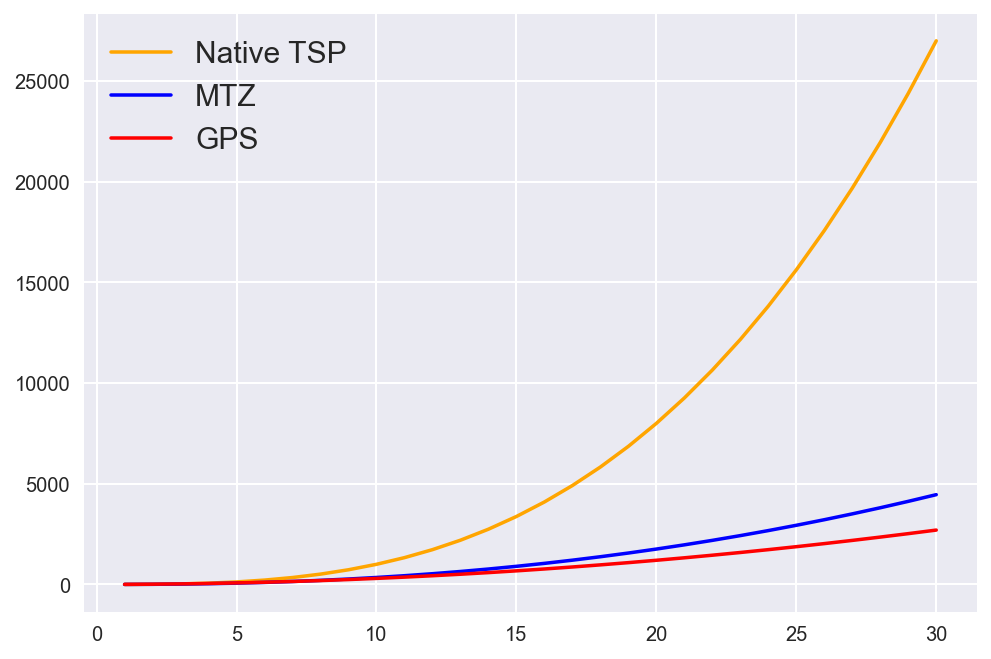}
	\caption{Comparison of the different models based on the number of qubits. This graph shows the behaviour and evolution of the numbers of qubits for each model. We see the best performance of our GPS model compared to the other models.}
	 \label{fig:GPSBench}
\end{figure}

\begin{figure}[ht!]
	\centering
	\includegraphics[width=0.4\textwidth]{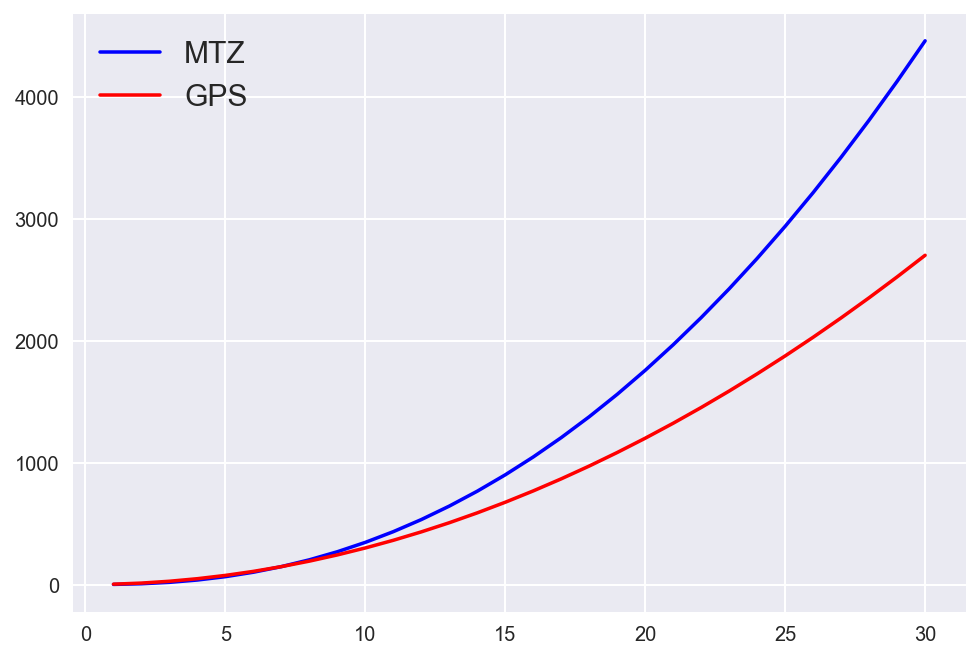}
	\caption{Benchmark between MTZ and GPS model based on the number of qubits. We can appreciate that for 30 cities, GPS model needs 2700 qubits while the MTZ 4458.}
	 \label{fig:MTZGPS}
\end{figure}

With this work, we offer a new formulation for the TSP called GPS and apply it to find an optimal formulation for the VRP that minimizes the time the vehicles make their journey. We have also seen that the results of the D-Wave simulator solver are consistent with the expected solution. However, we consider it unnecessary to test it in gate-based quantum computers, given their limitations today in the number of qubits. Still, we emphasize that our current formulation is valid for such computers. The improvement in our models represents a fairly significant order of magnitude because we went from $ N ^ 3 $ variables to $ 3N ^ 2 $. The figure \eqref{fig:GPSBench} and \eqref{fig:MTZGPS}  summarizes the major contribution of this article.

Our GPS formulation and the VRP proposal can help in optimization problems when we want to reduce the number of variables and therefore reduce the number of qubits quite a bit. In addition, it is interesting in situations, such as the one raised in the future line of the article \cite {a14070194}, by modelling some biological activities on selected sets of organic compounds as can be seen in \cite{modelingmolecular}, or resource optimization problems such as gasoline and aircraft travel. Another interesting application could be to compare GPS with the approach offered by this reference \cite{Zhang2020} using deep reinforcement learning to address combinatorial optimization problems with feasibility constraints. This leads us to project on how to make this comparison in quantum computing using the proposal made in this reference \cite{atchade2021quantum}.

The results obtained from our VRP formulation and all the experiments carried out maintain the number of variables $ QN ^ 2 $ and allow us to offer the community new formulations that minimize the time it takes for vehicles to travel. \\

Future work will apply the ideas developed in the QUBO model of these problems to similar ones. In particular, we will look for other variants of the TSP to use the modelling of this that we have carried out. \\

\textbf{Compliance with Ethics Guidelines}\\
Funding: This research received no external funding. 

Institutional review: This article does not contain any studies with human or animal subjects.

Informed consent: Informed consent was obtained from all individual participants included in the study.

Data availability: Data sharing not applicable. No new data were created or analyzed in this study. Data sharing is not applicable to this article.

\newpage

\section*{Appendix 1}\label{sec:Appendix}

\subsection{TSP formulation $N^2$.}\label{sec:TSPn2}
There is a TSP model that requires $ N ^ 2 $ variables, where these are the following:
\begin{equation}
    x_{i,t} \text{ such as } i\in \{0,...,N+1\} \text{ and } t\in \{0,...,N+1\}.
\end{equation}
Under this formulation $ x_ {i, t} = 1 $ denotes that the city $ i $ is reached at position $ t $. The distance calculation function with this formulation is as follows
\begin{equation}
    \sum_{i=0}^{N+1}\sum_{j=0}^{N+1}\sum_{t=0}^{N}d_{i,j}x_{i,t}x_{j,t+1},
\end{equation}
where $d_{i,j}$ represents the distance between the node $i$ and the node $j$. This expression has the problem that the distance formulation has terms of degree two and when trying to generalize this idea to other types of problems such as the VRP it will become a $4$ degree constraint making use of a large number of auxiliary variables to convert it to QUBO type format.

\subsection{Improved model $3N^2Q$}\label{sec:modelVRPenhanced}
In the previous modelling, we can improve the number of variables used from $ 5N ^ 2Q $ to $ 3N ^ 2Q $ since certain variables are redundant. Let us see how we can do this.
Let us take the set of variables

$$x_{i,j,r,q} \text{ with }i<j \in \{0,..,N+1\},r\in \{0,1,2\} \text{ and }q \in \{1,..,Q\} $$
In all the modelling, the variables $ x_ {i, j, r} $ such that $ i = j $ are not considered. Let us analyze the interpretation of each variable. For each edge $ (i, j) $, different cases depending on whether a vehicle passes through both cities, which city is visited before the other and whether the edge is travelled or not.
\begin{itemize}
    
    \item  $x_{i,j,0,q}=1$ means that the city $ i $ is reached earlier than the $ j $ and the edge $ (i, j) $ is not travelled.
    
    \item  $x_{i,j,1,q}=1$ means that the vehicle $ q $ travels the cities $ i $ and $ j $, it reaches the city $ i $ before the $ j $ and it travels the edge $ (i, j) $.
    
    \item  $x_{i,j,2,q}=1$ means that the city $ j $ is reached earlier than the $ i $ and the edge $ (j, i) $ is not travelled.
    
\end{itemize}

This new simplification keeps constraints \eqref{VRP:constrain2}, \eqref{VRP:constrain3},\eqref{VRP:constrain4}, \eqref{VRP:constrain5}, \eqref{QTSP-Constraint888} and \eqref{VRP:constrain11} defined in the same way as the first proposal of the VRP formulation, so we will only focus on the changes of the remaining constraints:

\begin{itemize}
    \item \textit{Constraint 1:} For each $i,j,q$, one and only one of the possibilities must be met for $ r $, so:
    \begin{equation}
      \text{For all }i,j,q \text{: } \sum_{r=0}^2 x_{i,j,r,q}=1,  
    \end{equation}
    
    \item\textit{Constraint 6:} That the city $ i $ is reached before the city $ j $ does not depend on each vehicle. Therefore, for all the vehicles that either arrive at city $ i $ earlier than $ j $, or arrive at city $ j $ earlier than $ i $. Introducing the auxiliary variables $ a_ {i, j} $, we have the following constraint. $ \text {For all} \quad i, j \in \{1, ..., N \} \text {:} $
    \begin{equation}
      \sum_{q=1}^Q x_{i,j,0,q}+x_{i,j,1,q} = a_{i,j} Q.  
      \label{rest72}
    \end{equation}

    \item\textit{Constraint 8:} It must be fulfilled that either the vehicle pass through the city $ i $ before the $ j $ or arrive before to the city $ j $ than the $ i $. Therefore, it must be verified that, $ \text {for} \quad i \in \{0, ..., N \} \text {,} \quad j \in \{1, ..., N \} \quad \text {and} \quad q \in \{1, ..., Q \} $:
    \begin{equation}
      x_{i,j,0,q}+x_{i,j,1,q}=1-(x_{j,i,0,q}+x_{j,i,1,q}).
    \end{equation}
    
    \item\textit{Constraint 9:} If the city $ i $ is reached before $ j $ and the city $ j $ is reached before the city $ k $, then the city $ i $ must be reached before the city $ k $. This condition will prevent the vehicle from returning to a city it has already passed through and therefore prevents a cycle from forming. \\ \begin{equation}
      \label{QTSP-Constraint102}
      \begin{aligned}
      & \lambda \sum_{i=1}^N\sum_{j=1}^N\sum_{k=1}^N (a_{i,j}a_{j,k}\\
      &- a_{i,j}a_{i,k} -a_{j,k}a_{i,k} +a_{i,k}^2),  
      \end{aligned}
    \end{equation}
    
\section{Restriction penalty \label{sec:GPSpenalization}}
 Let us analyze the system that must be solved to build the penalty function from the equation \eqref{GPS-Constraint5}. Our penalty function $ P (a_ {i, j}, a_ {j, k}, a_ {i, k}) $ must satisfy that $ P (0,0,1) = 1 $, $ P (1,1,0) = 1 $ and $ P (a_ {i, j}, a_ {j, k}, a_ {i, k}) = 0 $ for the rest of the cases. Let us call the variables $ a_ {i, j} = x $, $ a_ {j, k} = y $, $ a_ {i, k} = z $ to simplify the notation. Then, we arrive at the quadratic function $ P $, as is shown as follows:
 \begin{equation}
 \begin{aligned}
  & P(x,y,z) = c_1x^2+c_2xy+c_3xz+c_4y^2\\ 
  & +c_5yz +c_6z^2
 \end{aligned}
 \end{equation}
 
 Imposing the previous restrictions, we have the following system of equations.
 \begin{itemize}
     \item $P(0,0,1)=1$ So that $c_6 = 1$.
     \item $P(0,1,0)=0$ So that $c_4=0$.
     \item $P(0,1,1)=0$ So that $c_5+c_6 = 1 \Rightarrow c_5 = -1$.
     \item $P(1,0,0)=0$ So that $c_1 = 0$.
     \item $P(1,0,1)=0$ So that $c_1+c_3+c_6 = 0 \Rightarrow c_3 = -1$
     \item $P(1,1,0)=1$ So that $c_2 = 1$.\\
    So far, we have a system of 6 equations with six certain compatible unknowns. First, however, an additional restriction must be verified. Let us verify if it is met.
    \item $ P (1,1,1) = 0 $. $ \sum_{i = 1} ^ 6 c_i = 1-1-1 + 1 = 0 $. So that indeed all the requirements are met.

 \end{itemize}
 We then have that the following function which is a penalty function for the constraint \eqref{GPS-Constraint5}.
 \begin{equation}
 \begin{aligned}
  & P(a_{i,j}, a_{j,k}, a_{i,k}) = a_{i,j}a_{j,k} - a_{i,j}a_{i,k}\\
  & -a_{j,k}a_{i,k} +a_{i,k}^2
 \end{aligned}
 \end{equation}

\end{itemize}

\newpage
\bibliography{main}

\begin{thebibliography}{53}%
\makeatletter
\providecommand \@ifxundefined [1]{%
 \@ifx{#1\undefined}
}%
\providecommand \@ifnum [1]{%
 \ifnum #1\expandafter \@firstoftwo
 \else \expandafter \@secondoftwo
 \fi
}%
\providecommand \@ifx [1]{%
 \ifx #1\expandafter \@firstoftwo
 \else \expandafter \@secondoftwo
 \fi
}%
\providecommand \natexlab [1]{#1}%
\providecommand \enquote  [1]{``#1''}%
\providecommand \bibnamefont  [1]{#1}%
\providecommand \bibfnamefont [1]{#1}%
\providecommand \citenamefont [1]{#1}%
\providecommand \href@noop [0]{\@secondoftwo}%
\providecommand \href [0]{\begingroup \@sanitize@url \@href}%
\providecommand \@href[1]{\@@startlink{#1}\@@href}%
\providecommand \@@href[1]{\endgroup#1\@@endlink}%
\providecommand \@sanitize@url [0]{\catcode `\\12\catcode `\$12\catcode
  `\&12\catcode `\#12\catcode `\^12\catcode `\_12\catcode `\%12\relax}%
\providecommand \@@startlink[1]{}%
\providecommand \@@endlink[0]{}%
\providecommand \url  [0]{\begingroup\@sanitize@url \@url }%
\providecommand \@url [1]{\endgroup\@href {#1}{\urlprefix }}%
\providecommand \urlprefix  [0]{URL }%
\providecommand \Eprint [0]{\href }%
\providecommand \doibase [0]{https://doi.org/}%
\providecommand \selectlanguage [0]{\@gobble}%
\providecommand \bibinfo  [0]{\@secondoftwo}%
\providecommand \bibfield  [0]{\@secondoftwo}%
\providecommand \translation [1]{[#1]}%
\providecommand \BibitemOpen [0]{}%
\providecommand \bibitemStop [0]{}%
\providecommand \bibitemNoStop [0]{.\EOS\space}%
\providecommand \EOS [0]{\spacefactor3000\relax}%
\providecommand \BibitemShut  [1]{\csname bibitem#1\endcsname}%
\let\auto@bib@innerbib\@empty
\bibitem [{\citenamefont {Gavish}\ and\ \citenamefont
  {Graves}(1978)}]{gavish1978travelling}%
  \BibitemOpen
  \bibfield  {author} {\bibinfo {author} {\bibfnamefont {B.}~\bibnamefont
  {Gavish}}\ and\ \bibinfo {author} {\bibfnamefont {S.~C.}\ \bibnamefont
  {Graves}},\ }\bibfield  {title} {\bibinfo {title} {The travelling salesman
  problem and related problems},\ }\href@noop {} {\bibfield  {journal}
  {\bibinfo  {journal} {MIT}\ } (\bibinfo {year} {1978})}\BibitemShut {NoStop}%
\bibitem [{\citenamefont {Hochba}(1997)}]{hochba1997approximation}%
  \BibitemOpen
  \bibfield  {author} {\bibinfo {author} {\bibfnamefont {D.~S.}\ \bibnamefont
  {Hochba}},\ }\bibfield  {title} {\bibinfo {title} {Approximation algorithms
  for np-hard problems},\ }\href@noop {} {\bibfield  {journal} {\bibinfo
  {journal} {ACM Sigact News}\ }\textbf {\bibinfo {volume} {28}},\ \bibinfo
  {pages} {40} (\bibinfo {year} {1997})}\BibitemShut {NoStop}%
\bibitem [{\citenamefont {Lewis}\ and\ \citenamefont
  {Glover}(2017)}]{lewis2017quadratic}%
  \BibitemOpen
  \bibfield  {author} {\bibinfo {author} {\bibfnamefont {M.}~\bibnamefont
  {Lewis}}\ and\ \bibinfo {author} {\bibfnamefont {F.}~\bibnamefont {Glover}},\
  }\bibfield  {title} {\bibinfo {title} {Quadratic unconstrained binary
  optimization problem preprocessing: Theory and empirical analysis},\
  }\href@noop {} {\bibfield  {journal} {\bibinfo  {journal} {Networks}\
  }\textbf {\bibinfo {volume} {70}},\ \bibinfo {pages} {79} (\bibinfo {year}
  {2017})}\BibitemShut {NoStop}%
\bibitem [{\citenamefont {Cipra}(1987)}]{cipra1987introduction}%
  \BibitemOpen
  \bibfield  {author} {\bibinfo {author} {\bibfnamefont {B.~A.}\ \bibnamefont
  {Cipra}},\ }\bibfield  {title} {\bibinfo {title} {An introduction to the
  ising model},\ }\href@noop {} {\bibfield  {journal} {\bibinfo  {journal} {The
  American Mathematical Monthly}\ }\textbf {\bibinfo {volume} {94}},\ \bibinfo
  {pages} {937} (\bibinfo {year} {1987})}\BibitemShut {NoStop}%
\bibitem [{\citenamefont {McGeoch}\ and\ \citenamefont
  {Wang}(2013)}]{McGeoch2013}%
  \BibitemOpen
  \bibfield  {author} {\bibinfo {author} {\bibfnamefont {C.~C.}\ \bibnamefont
  {McGeoch}}\ and\ \bibinfo {author} {\bibfnamefont {C.}~\bibnamefont {Wang}},\
  }\bibfield  {title} {\bibinfo {title} {Experimental evaluation of an
  adiabiatic quantum system for combinatorial optimization},\ }in\ \href
  {https://doi.org/10.1145/2482767.2482797} {\emph {\bibinfo {booktitle}
  {Proceedings of the {ACM} International Conference on Computing Frontiers -
  {CF} {\textquotesingle}13}}}\ (\bibinfo  {publisher} {{ACM} Press},\ \bibinfo
  {year} {2013})\BibitemShut {NoStop}%
\bibitem [{\citenamefont {McGeoch}(2014)}]{mcgeoch2014adiabatic}%
  \BibitemOpen
  \bibfield  {author} {\bibinfo {author} {\bibfnamefont {C.~C.}\ \bibnamefont
  {McGeoch}},\ }\bibfield  {title} {\bibinfo {title} {Adiabatic quantum
  computation and quantum annealing: Theory and practice},\ }\href@noop {}
  {\bibfield  {journal} {\bibinfo  {journal} {Synthesis Lectures on Quantum
  Computing}\ }\textbf {\bibinfo {volume} {5}},\ \bibinfo {pages} {1} (\bibinfo
  {year} {2014})}\BibitemShut {NoStop}%
\bibitem [{\citenamefont {Brooke}\ \emph {et~al.}(1999)\citenamefont {Brooke},
  \citenamefont {Bitko}, \citenamefont {Rosenbaum},\ and\ \citenamefont
  {Aeppli}}]{brooke1999quantum}%
  \BibitemOpen
  \bibfield  {author} {\bibinfo {author} {\bibfnamefont {J.}~\bibnamefont
  {Brooke}}, \bibinfo {author} {\bibfnamefont {D.}~\bibnamefont {Bitko}},
  \bibinfo {author} {\bibnamefont {Rosenbaum}},\ and\ \bibinfo {author}
  {\bibfnamefont {G.}~\bibnamefont {Aeppli}},\ }\bibfield  {title} {\bibinfo
  {title} {Quantum annealing of a disordered magnet},\ }\href@noop {}
  {\bibfield  {journal} {\bibinfo  {journal} {Science}\ }\textbf {\bibinfo
  {volume} {284}},\ \bibinfo {pages} {779} (\bibinfo {year}
  {1999})}\BibitemShut {NoStop}%
\bibitem [{\citenamefont {Born}\ and\ \citenamefont {Fock}(1928)}]{Born1928}%
  \BibitemOpen
  \bibfield  {author} {\bibinfo {author} {\bibfnamefont {M.}~\bibnamefont
  {Born}}\ and\ \bibinfo {author} {\bibfnamefont {V.}~\bibnamefont {Fock}},\
  }\bibfield  {title} {\bibinfo {title} {Beweis des adiabatensatzes},\ }\href
  {https://doi.org/10.1007/bf01343193} {\bibfield  {journal} {\bibinfo
  {journal} {Zeitschrift für Physik}\ }\textbf {\bibinfo {volume} {51}},\
  \bibinfo {pages} {165} (\bibinfo {year} {1928})}\BibitemShut {NoStop}%
\bibitem [{\citenamefont {Farhi}\ \emph {et~al.}(2000)\citenamefont {Farhi},
  \citenamefont {Goldstone}, \citenamefont {Gutmann},\ and\ \citenamefont
  {Sipser}}]{farhi2000quantum}%
  \BibitemOpen
  \bibfield  {author} {\bibinfo {author} {\bibfnamefont {E.}~\bibnamefont
  {Farhi}}, \bibinfo {author} {\bibfnamefont {J.}~\bibnamefont {Goldstone}},
  \bibinfo {author} {\bibfnamefont {S.}~\bibnamefont {Gutmann}},\ and\ \bibinfo
  {author} {\bibfnamefont {M.}~\bibnamefont {Sipser}},\ }\bibfield  {title}
  {\bibinfo {title} {Quantum computation by adiabatic evolution},\ }\href@noop
  {} {\bibfield  {journal} {\bibinfo  {journal} {arXiv preprint
  quant-ph/0001106}\ } (\bibinfo {year} {2000})}\BibitemShut {NoStop}%
\bibitem [{\citenamefont {Bian}\ \emph {et~al.}(2010)\citenamefont {Bian},
  \citenamefont {Chudak}, \citenamefont {Macready},\ and\ \citenamefont
  {Rose}}]{bian2010ising}%
  \BibitemOpen
  \bibfield  {author} {\bibinfo {author} {\bibfnamefont {Z.}~\bibnamefont
  {Bian}}, \bibinfo {author} {\bibfnamefont {F.}~\bibnamefont {Chudak}},
  \bibinfo {author} {\bibfnamefont {W.~G.}\ \bibnamefont {Macready}},\ and\
  \bibinfo {author} {\bibfnamefont {G.}~\bibnamefont {Rose}},\ }\bibfield
  {title} {\bibinfo {title} {The ising model: teaching an old problem new
  tricks},\ }\href@noop {} {\bibfield  {journal} {\bibinfo  {journal} {D-wave
  systems}\ }\textbf {\bibinfo {volume} {2}} (\bibinfo {year}
  {2010})}\BibitemShut {NoStop}%
\bibitem [{\citenamefont {Toth}\ and\ \citenamefont
  {Vigo}(2002{\natexlab{a}})}]{toth2002vehicle}%
  \BibitemOpen
  \bibfield  {author} {\bibinfo {author} {\bibfnamefont {P.}~\bibnamefont
  {Toth}}\ and\ \bibinfo {author} {\bibfnamefont {D.}~\bibnamefont {Vigo}},\
  }\href@noop {} {\emph {\bibinfo {title} {The vehicle routing problem}}}\
  (\bibinfo  {publisher} {SIAM},\ \bibinfo {year} {2002})\BibitemShut {NoStop}%
\bibitem [{\citenamefont {Toth}\ and\ \citenamefont
  {Vigo}(2002{\natexlab{b}})}]{toth2002models}%
  \BibitemOpen
  \bibfield  {author} {\bibinfo {author} {\bibfnamefont {P.}~\bibnamefont
  {Toth}}\ and\ \bibinfo {author} {\bibfnamefont {D.}~\bibnamefont {Vigo}},\
  }\bibfield  {title} {\bibinfo {title} {Models, relaxations and exact
  approaches for the capacitated vehicle routing problem},\ }\href@noop {}
  {\bibfield  {journal} {\bibinfo  {journal} {Discrete Applied Mathematics}\
  }\textbf {\bibinfo {volume} {123}},\ \bibinfo {pages} {487} (\bibinfo {year}
  {2002}{\natexlab{b}})}\BibitemShut {NoStop}%
\bibitem [{\citenamefont {Ralphs}\ \emph {et~al.}(2003)\citenamefont {Ralphs},
  \citenamefont {Kopman}, \citenamefont {Pulleyblank},\ and\ \citenamefont
  {Trotter}}]{ralphs2003capacitated}%
  \BibitemOpen
  \bibfield  {author} {\bibinfo {author} {\bibfnamefont {T.~K.}\ \bibnamefont
  {Ralphs}}, \bibinfo {author} {\bibfnamefont {L.}~\bibnamefont {Kopman}},
  \bibinfo {author} {\bibfnamefont {W.~R.}\ \bibnamefont {Pulleyblank}},\ and\
  \bibinfo {author} {\bibfnamefont {L.~E.}\ \bibnamefont {Trotter}},\
  }\bibfield  {title} {\bibinfo {title} {On the capacitated vehicle routing
  problem},\ }\href@noop {} {\bibfield  {journal} {\bibinfo  {journal}
  {Mathematical programming}\ }\textbf {\bibinfo {volume} {94}},\ \bibinfo
  {pages} {343} (\bibinfo {year} {2003})}\BibitemShut {NoStop}%
\bibitem [{\citenamefont {Atchade-Adelomou}\ \emph
  {et~al.}(2021{\natexlab{a}})\citenamefont {Atchade-Adelomou}, \citenamefont
  {Alonso-Linaje}, \citenamefont {Albo-Canals},\ and\ \citenamefont
  {Casado-Fauli}}]{a14070194}%
  \BibitemOpen
  \bibfield  {author} {\bibinfo {author} {\bibfnamefont {P.}~\bibnamefont
  {Atchade-Adelomou}}, \bibinfo {author} {\bibfnamefont {G.}~\bibnamefont
  {Alonso-Linaje}}, \bibinfo {author} {\bibfnamefont {J.}~\bibnamefont
  {Albo-Canals}},\ and\ \bibinfo {author} {\bibfnamefont {D.}~\bibnamefont
  {Casado-Fauli}},\ }\bibfield  {title} {\bibinfo {title} {qrobot: A quantum
  computing approach in mobile robot order picking and batching problem solver
  optimization},\ }\bibfield  {journal} {\bibinfo  {journal} {Algorithms}\
  }\textbf {\bibinfo {volume} {14}},\ \href {https://doi.org/10.3390/a14070194}
  {10.3390/a14070194} (\bibinfo {year} {2021}{\natexlab{a}})\BibitemShut
  {NoStop}%
\bibitem [{\citenamefont {Miller}\ \emph {et~al.}(1960)\citenamefont {Miller},
  \citenamefont {Tucker},\ and\ \citenamefont {Zemlin}}]{miller1960integer}%
  \BibitemOpen
  \bibfield  {author} {\bibinfo {author} {\bibfnamefont {C.~E.}\ \bibnamefont
  {Miller}}, \bibinfo {author} {\bibfnamefont {A.~W.}\ \bibnamefont {Tucker}},\
  and\ \bibinfo {author} {\bibfnamefont {R.~A.}\ \bibnamefont {Zemlin}},\
  }\bibfield  {title} {\bibinfo {title} {Integer programming formulation of
  traveling salesman problems},\ }\href@noop {} {\bibfield  {journal} {\bibinfo
   {journal} {Journal of the ACM (JACM)}\ }\textbf {\bibinfo {volume} {7}},\
  \bibinfo {pages} {326} (\bibinfo {year} {1960})}\BibitemShut {NoStop}%
\bibitem [{\citenamefont {Boruvka}(1926{\natexlab{a}})}]{boruvka1926minimal}%
  \BibitemOpen
  \bibfield  {author} {\bibinfo {author} {\bibfnamefont {O.}~\bibnamefont
  {Boruvka}},\ }\bibfield  {title} {\bibinfo {title} {On a minimal problem},\
  }\href@noop {} {\bibfield  {journal} {\bibinfo  {journal} {Pr{\'a}ce
  Moravsk{\'e} Pridovedeck{\'e} Spolecnosti}\ }\textbf {\bibinfo {volume}
  {3}},\ \bibinfo {pages} {37} (\bibinfo {year}
  {1926}{\natexlab{a}})}\BibitemShut {NoStop}%
\bibitem [{\citenamefont {Boruvka}(1926{\natexlab{b}})}]{boruvka1926prispevek}%
  \BibitemOpen
  \bibfield  {author} {\bibinfo {author} {\bibfnamefont {O.}~\bibnamefont
  {Boruvka}},\ }\bibfield  {title} {\bibinfo {title} {Pr{\i}spevek k
  re{\v{s}}en{\i} ot{\'a}zky ekonomick{\'e} stavby elektrovodn{\i}ch s{\i}t{\i}
  (contribution to the solution of a problem of economical construction of
  electrical networks)},\ }\href@noop {} {\bibfield  {journal} {\bibinfo
  {journal} {Elektronick{\`y} obzor}\ }\textbf {\bibinfo {volume} {15}},\
  \bibinfo {pages} {153} (\bibinfo {year} {1926}{\natexlab{b}})}\BibitemShut
  {NoStop}%
\bibitem [{\citenamefont {Kruskal}(1956)}]{kruskal1956shortest}%
  \BibitemOpen
  \bibfield  {author} {\bibinfo {author} {\bibfnamefont {J.~B.}\ \bibnamefont
  {Kruskal}},\ }\bibfield  {title} {\bibinfo {title} {On the shortest spanning
  subtree of a graph and the traveling salesman problem},\ }\href@noop {}
  {\bibfield  {journal} {\bibinfo  {journal} {Proceedings of the American
  Mathematical society}\ }\textbf {\bibinfo {volume} {7}},\ \bibinfo {pages}
  {48} (\bibinfo {year} {1956})}\BibitemShut {NoStop}%
\bibitem [{\citenamefont {Bellmore}\ and\ \citenamefont
  {Nemhauser}(1968)}]{Bellmore1968}%
  \BibitemOpen
  \bibfield  {author} {\bibinfo {author} {\bibfnamefont {M.}~\bibnamefont
  {Bellmore}}\ and\ \bibinfo {author} {\bibfnamefont {G.~L.}\ \bibnamefont
  {Nemhauser}},\ }\bibfield  {title} {\bibinfo {title} {The traveling salesman
  problem: A survey},\ }\href {https://doi.org/10.1287/opre.16.3.538}
  {\bibfield  {journal} {\bibinfo  {journal} {Operations Research}\ }\textbf
  {\bibinfo {volume} {16}},\ \bibinfo {pages} {538} (\bibinfo {year}
  {1968})}\BibitemShut {NoStop}%
\bibitem [{\citenamefont {Lenstra}\ and\ \citenamefont
  {Kan}(1975)}]{Lenstra1975}%
  \BibitemOpen
  \bibfield  {author} {\bibinfo {author} {\bibfnamefont {J.~K.}\ \bibnamefont
  {Lenstra}}\ and\ \bibinfo {author} {\bibfnamefont {A.~H. G.~R.}\ \bibnamefont
  {Kan}},\ }\bibfield  {title} {\bibinfo {title} {Some simple applications of
  the travelling salesman problem},\ }\href
  {https://doi.org/10.1057/jors.1975.151} {\bibfield  {journal} {\bibinfo
  {journal} {Journal of the Operational Research Society}\ }\textbf {\bibinfo
  {volume} {26}},\ \bibinfo {pages} {717} (\bibinfo {year} {1975})}\BibitemShut
  {NoStop}%
\bibitem [{\citenamefont {Adelomou}\ \emph {et~al.}(2020)\citenamefont
  {Adelomou}, \citenamefont {Rib{\'e}},\ and\ \citenamefont
  {Cardona}}]{adelomou2020formulation}%
  \BibitemOpen
  \bibfield  {author} {\bibinfo {author} {\bibfnamefont {A.~P.}\ \bibnamefont
  {Adelomou}}, \bibinfo {author} {\bibfnamefont {E.~G.}\ \bibnamefont
  {Rib{\'e}}},\ and\ \bibinfo {author} {\bibfnamefont {X.~V.}\ \bibnamefont
  {Cardona}},\ }\bibfield  {title} {\bibinfo {title} {Formulation of the social
  workers’ problem in quadratic unconstrained binary optimization form and
  solve it on a quantum computer},\ }\href@noop {} {\bibfield  {journal}
  {\bibinfo  {journal} {Journal of Computer and Communications}\ }\textbf
  {\bibinfo {volume} {8}},\ \bibinfo {pages} {44} (\bibinfo {year}
  {2020})}\BibitemShut {NoStop}%
\bibitem [{\citenamefont {Lee}\ \emph {et~al.}(2004)\citenamefont {Lee},
  \citenamefont {Shin}, \citenamefont {Park},\ and\ \citenamefont
  {Zhang}}]{Lee2004}%
  \BibitemOpen
  \bibfield  {author} {\bibinfo {author} {\bibfnamefont {J.~Y.}\ \bibnamefont
  {Lee}}, \bibinfo {author} {\bibfnamefont {S.-Y.}\ \bibnamefont {Shin}},
  \bibinfo {author} {\bibfnamefont {T.~H.}\ \bibnamefont {Park}},\ and\
  \bibinfo {author} {\bibfnamefont {B.-T.}\ \bibnamefont {Zhang}},\ }\bibfield
  {title} {\bibinfo {title} {Solving traveling salesman problems with {DNA}
  molecules encoding numerical values},\ }\href
  {https://doi.org/10.1016/j.biosystems.2004.06.005} {\bibfield  {journal}
  {\bibinfo  {journal} {Biosystems}\ }\textbf {\bibinfo {volume} {78}},\
  \bibinfo {pages} {39} (\bibinfo {year} {2004})}\BibitemShut {NoStop}%
\bibitem [{\citenamefont {Ball}(2000)}]{Ball2000}%
  \BibitemOpen
  \bibfield  {author} {\bibinfo {author} {\bibfnamefont {P.}~\bibnamefont
  {Ball}},\ }\bibfield  {title} {\bibinfo {title} {{DNA} computer helps
  travelling salesman},\ }\bibfield  {journal} {\bibinfo  {journal} {Nature}\
  }\href {https://doi.org/10.1038/news000113-10} {10.1038/news000113-10}
  (\bibinfo {year} {2000})\BibitemShut {NoStop}%
\bibitem [{\citenamefont {Pataki}(2003)}]{pataki2003teaching}%
  \BibitemOpen
  \bibfield  {author} {\bibinfo {author} {\bibfnamefont {G.}~\bibnamefont
  {Pataki}},\ }\bibfield  {title} {\bibinfo {title} {Teaching integer
  programming formulations using the traveling salesman problem},\ }\href@noop
  {} {\bibfield  {journal} {\bibinfo  {journal} {SIAM review}\ }\textbf
  {\bibinfo {volume} {45}},\ \bibinfo {pages} {116} (\bibinfo {year}
  {2003})}\BibitemShut {NoStop}%
\bibitem [{\citenamefont {Marto{\v{n}}{\'a}k}\ \emph
  {et~al.}(2004)\citenamefont {Marto{\v{n}}{\'a}k}, \citenamefont {Santoro},\
  and\ \citenamefont {Tosatti}}]{martovnak2004quantum}%
  \BibitemOpen
  \bibfield  {author} {\bibinfo {author} {\bibfnamefont {R.}~\bibnamefont
  {Marto{\v{n}}{\'a}k}}, \bibinfo {author} {\bibfnamefont {G.~E.}\ \bibnamefont
  {Santoro}},\ and\ \bibinfo {author} {\bibfnamefont {E.}~\bibnamefont
  {Tosatti}},\ }\bibfield  {title} {\bibinfo {title} {Quantum annealing of the
  traveling-salesman problem},\ }\href@noop {} {\bibfield  {journal} {\bibinfo
  {journal} {Physical Review E}\ }\textbf {\bibinfo {volume} {70}},\ \bibinfo
  {pages} {057701} (\bibinfo {year} {2004})}\BibitemShut {NoStop}%
\bibitem [{\citenamefont {Warren}(2013)}]{warren2013adapting}%
  \BibitemOpen
  \bibfield  {author} {\bibinfo {author} {\bibfnamefont {R.~H.}\ \bibnamefont
  {Warren}},\ }\bibfield  {title} {\bibinfo {title} {Adapting the traveling
  salesman problem to an adiabatic quantum computer},\ }\href@noop {}
  {\bibfield  {journal} {\bibinfo  {journal} {Quantum information processing}\
  }\textbf {\bibinfo {volume} {12}},\ \bibinfo {pages} {1781} (\bibinfo {year}
  {2013})}\BibitemShut {NoStop}%
\bibitem [{\citenamefont {Warren}(2017)}]{warren2017small}%
  \BibitemOpen
  \bibfield  {author} {\bibinfo {author} {\bibfnamefont {R.~H.}\ \bibnamefont
  {Warren}},\ }\bibfield  {title} {\bibinfo {title} {Small traveling salesman
  problems},\ }\href@noop {} {\bibfield  {journal} {\bibinfo  {journal}
  {Journal of Advances in Applied Mathematics}\ }\textbf {\bibinfo {volume}
  {2}} (\bibinfo {year} {2017})}\BibitemShut {NoStop}%
\bibitem [{\citenamefont {Greco}(2008)}]{greco2008traveling}%
  \BibitemOpen
  \bibfield  {author} {\bibinfo {author} {\bibfnamefont {F.}~\bibnamefont
  {Greco}},\ }\href@noop {} {\emph {\bibinfo {title} {Traveling salesman
  problem}}}\ (\bibinfo  {publisher} {BoD--Books on Demand},\ \bibinfo {year}
  {2008})\BibitemShut {NoStop}%
\bibitem [{\citenamefont {Reinelt}(1991)}]{reinelt1991tsplib}%
  \BibitemOpen
  \bibfield  {author} {\bibinfo {author} {\bibfnamefont {G.}~\bibnamefont
  {Reinelt}},\ }\bibfield  {title} {\bibinfo {title} {Tsplib—a traveling
  salesman problem library},\ }\href@noop {} {\bibfield  {journal} {\bibinfo
  {journal} {ORSA journal on computing}\ }\textbf {\bibinfo {volume} {3}},\
  \bibinfo {pages} {376} (\bibinfo {year} {1991})}\BibitemShut {NoStop}%
\bibitem [{\citenamefont {Crosson}\ and\ \citenamefont
  {Harrow}(2016)}]{crosson2016simulated}%
  \BibitemOpen
  \bibfield  {author} {\bibinfo {author} {\bibfnamefont {E.}~\bibnamefont
  {Crosson}}\ and\ \bibinfo {author} {\bibfnamefont {A.~W.}\ \bibnamefont
  {Harrow}},\ }\bibfield  {title} {\bibinfo {title} {Simulated quantum
  annealing can be exponentially faster than classical simulated annealing},\
  }in\ \href@noop {} {\emph {\bibinfo {booktitle} {2016 IEEE 57th Annual
  Symposium on Foundations of Computer Science (FOCS)}}}\ (\bibinfo
  {organization} {IEEE},\ \bibinfo {year} {2016})\ pp.\ \bibinfo {pages}
  {714--723}\BibitemShut {NoStop}%
\bibitem [{\citenamefont {Leung}\ \emph {et~al.}(2004)\citenamefont {Leung},
  \citenamefont {Jin},\ and\ \citenamefont {Xu}}]{leung2004expanding}%
  \BibitemOpen
  \bibfield  {author} {\bibinfo {author} {\bibfnamefont {K.-S.}\ \bibnamefont
  {Leung}}, \bibinfo {author} {\bibfnamefont {H.-D.}\ \bibnamefont {Jin}},\
  and\ \bibinfo {author} {\bibfnamefont {Z.-B.}\ \bibnamefont {Xu}},\
  }\bibfield  {title} {\bibinfo {title} {An expanding self-organizing neural
  network for the traveling salesman problem},\ }\href@noop {} {\bibfield
  {journal} {\bibinfo  {journal} {Neurocomputing}\ }\textbf {\bibinfo {volume}
  {62}},\ \bibinfo {pages} {267} (\bibinfo {year} {2004})}\BibitemShut
  {NoStop}%
\bibitem [{\citenamefont {Cochrane}\ and\ \citenamefont
  {Beasley}(2003)}]{cochrane2003co}%
  \BibitemOpen
  \bibfield  {author} {\bibinfo {author} {\bibfnamefont {E.}~\bibnamefont
  {Cochrane}}\ and\ \bibinfo {author} {\bibfnamefont {J.~E.}\ \bibnamefont
  {Beasley}},\ }\bibfield  {title} {\bibinfo {title} {The co-adaptive neural
  network approach to the euclidean travelling salesman problem},\ }\href@noop
  {} {\bibfield  {journal} {\bibinfo  {journal} {Neural Networks}\ }\textbf
  {\bibinfo {volume} {16}},\ \bibinfo {pages} {1499} (\bibinfo {year}
  {2003})}\BibitemShut {NoStop}%
\bibitem [{\citenamefont {Feld}\ \emph {et~al.}(2019)\citenamefont {Feld},
  \citenamefont {Roch}, \citenamefont {Gabor}, \citenamefont {Seidel},
  \citenamefont {Neukart}, \citenamefont {Galter}, \citenamefont {Mauerer},\
  and\ \citenamefont {Linnhoff-Popien}}]{feld2019hybrid}%
  \BibitemOpen
  \bibfield  {author} {\bibinfo {author} {\bibfnamefont {S.}~\bibnamefont
  {Feld}}, \bibinfo {author} {\bibfnamefont {C.}~\bibnamefont {Roch}}, \bibinfo
  {author} {\bibfnamefont {T.}~\bibnamefont {Gabor}}, \bibinfo {author}
  {\bibfnamefont {C.}~\bibnamefont {Seidel}}, \bibinfo {author} {\bibfnamefont
  {F.}~\bibnamefont {Neukart}}, \bibinfo {author} {\bibfnamefont
  {I.}~\bibnamefont {Galter}}, \bibinfo {author} {\bibfnamefont
  {W.}~\bibnamefont {Mauerer}},\ and\ \bibinfo {author} {\bibfnamefont
  {C.}~\bibnamefont {Linnhoff-Popien}},\ }\bibfield  {title} {\bibinfo {title}
  {A hybrid solution method for the capacitated vehicle routing problem using a
  quantum annealer},\ }\href@noop {} {\bibfield  {journal} {\bibinfo  {journal}
  {Frontiers in ICT}\ }\textbf {\bibinfo {volume} {6}},\ \bibinfo {pages} {13}
  (\bibinfo {year} {2019})}\BibitemShut {NoStop}%
\bibitem [{\citenamefont {Irie}\ \emph {et~al.}(2019)\citenamefont {Irie},
  \citenamefont {Wongpaisarnsin}, \citenamefont {Terabe}, \citenamefont
  {Miki},\ and\ \citenamefont {Taguchi}}]{irie2019quantum}%
  \BibitemOpen
  \bibfield  {author} {\bibinfo {author} {\bibfnamefont {H.}~\bibnamefont
  {Irie}}, \bibinfo {author} {\bibfnamefont {G.}~\bibnamefont
  {Wongpaisarnsin}}, \bibinfo {author} {\bibfnamefont {M.}~\bibnamefont
  {Terabe}}, \bibinfo {author} {\bibfnamefont {A.}~\bibnamefont {Miki}},\ and\
  \bibinfo {author} {\bibfnamefont {S.}~\bibnamefont {Taguchi}},\ }\bibfield
  {title} {\bibinfo {title} {Quantum annealing of vehicle routing problem with
  time, state and capacity},\ }in\ \href@noop {} {\emph {\bibinfo {booktitle}
  {International Workshop on Quantum Technology and Optimization Problems}}}\
  (\bibinfo {organization} {Springer},\ \bibinfo {year} {2019})\ pp.\ \bibinfo
  {pages} {145--156}\BibitemShut {NoStop}%
\bibitem [{\citenamefont {Focacci}\ \emph {et~al.}(2002)\citenamefont
  {Focacci}, \citenamefont {Lodi},\ and\ \citenamefont
  {Milano}}]{focacci2002hybrid}%
  \BibitemOpen
  \bibfield  {author} {\bibinfo {author} {\bibfnamefont {F.}~\bibnamefont
  {Focacci}}, \bibinfo {author} {\bibfnamefont {A.}~\bibnamefont {Lodi}},\ and\
  \bibinfo {author} {\bibfnamefont {M.}~\bibnamefont {Milano}},\ }\bibfield
  {title} {\bibinfo {title} {A hybrid exact algorithm for the tsptw},\
  }\href@noop {} {\bibfield  {journal} {\bibinfo  {journal} {INFORMS journal on
  Computing}\ }\textbf {\bibinfo {volume} {14}},\ \bibinfo {pages} {403}
  (\bibinfo {year} {2002})}\BibitemShut {NoStop}%
\bibitem [{\citenamefont {Edelkamp}\ \emph {et~al.}(2013)\citenamefont
  {Edelkamp}, \citenamefont {Gath}, \citenamefont {Cazenave},\ and\
  \citenamefont {Teytaud}}]{edelkamp2013algorithm}%
  \BibitemOpen
  \bibfield  {author} {\bibinfo {author} {\bibfnamefont {S.}~\bibnamefont
  {Edelkamp}}, \bibinfo {author} {\bibfnamefont {M.}~\bibnamefont {Gath}},
  \bibinfo {author} {\bibfnamefont {T.}~\bibnamefont {Cazenave}},\ and\
  \bibinfo {author} {\bibfnamefont {F.}~\bibnamefont {Teytaud}},\ }\bibfield
  {title} {\bibinfo {title} {Algorithm and knowledge engineering for the tsptw
  problem},\ }in\ \href@noop {} {\emph {\bibinfo {booktitle} {2013 IEEE
  Symposium on Computational Intelligence in Scheduling (CISched)}}}\ (\bibinfo
  {organization} {IEEE},\ \bibinfo {year} {2013})\ pp.\ \bibinfo {pages}
  {44--51}\BibitemShut {NoStop}%
\bibitem [{\citenamefont {Atchade-Adelomou}\ \emph
  {et~al.}(2020{\natexlab{a}})\citenamefont {Atchade-Adelomou}, \citenamefont
  {Golobardes-Rib{\'{e}}},\ and\ \citenamefont
  {Vilas{\'{\i}}s-Cardona}}]{AtchadeAdelomou2020}%
  \BibitemOpen
  \bibfield  {author} {\bibinfo {author} {\bibfnamefont {P.}~\bibnamefont
  {Atchade-Adelomou}}, \bibinfo {author} {\bibfnamefont {E.}~\bibnamefont
  {Golobardes-Rib{\'{e}}}},\ and\ \bibinfo {author} {\bibfnamefont
  {X.}~\bibnamefont {Vilas{\'{\i}}s-Cardona}},\ }\bibfield  {title} {\bibinfo
  {title} {Using the variational-quantum-eigensolver ({VQE}) to create an
  intelligent social workers schedule problem solver},\ }in\ \href
  {https://doi.org/10.1007/978-3-030-61705-9_21} {\emph {\bibinfo {booktitle}
  {Lecture Notes in Computer Science}}}\ (\bibinfo  {publisher} {Springer
  International Publishing},\ \bibinfo {year} {2020})\ pp.\ \bibinfo {pages}
  {245--260}\BibitemShut {NoStop}%
\bibitem [{\citenamefont {Atchade-Adelomou}\ \emph
  {et~al.}(2020{\natexlab{b}})\citenamefont {Atchade-Adelomou}, \citenamefont
  {Golobardes-Ribe},\ and\ \citenamefont
  {Vilasis-Cardona}}]{adelomou2020using}%
  \BibitemOpen
  \bibfield  {author} {\bibinfo {author} {\bibfnamefont {P.}~\bibnamefont
  {Atchade-Adelomou}}, \bibinfo {author} {\bibfnamefont {E.}~\bibnamefont
  {Golobardes-Ribe}},\ and\ \bibinfo {author} {\bibfnamefont {X.}~\bibnamefont
  {Vilasis-Cardona}},\ }\href@noop {} {\bibinfo {title} {Using the
  parameterized quantum circuit combined with variational-quantum-eigensolver
  (vqe) to create an intelligent social workers' schedule problem solver}}
  (\bibinfo {year} {2020}{\natexlab{b}}),\ \Eprint
  {https://arxiv.org/abs/2010.05863} {arXiv:2010.05863 [quant-ph]} \BibitemShut
  {NoStop}%
\bibitem [{\citenamefont {Atchade-Adelomou}\ \emph
  {et~al.}(2021{\natexlab{b}})\citenamefont {Atchade-Adelomou}, \citenamefont
  {Casado-Fauli}, \citenamefont {Golobardes-Ribe},\ and\ \citenamefont
  {Vilasis-Cardona}}]{atchadeadelomou2021quantum}%
  \BibitemOpen
  \bibfield  {author} {\bibinfo {author} {\bibfnamefont {P.}~\bibnamefont
  {Atchade-Adelomou}}, \bibinfo {author} {\bibfnamefont {D.}~\bibnamefont
  {Casado-Fauli}}, \bibinfo {author} {\bibfnamefont {E.}~\bibnamefont
  {Golobardes-Ribe}},\ and\ \bibinfo {author} {\bibfnamefont {X.}~\bibnamefont
  {Vilasis-Cardona}},\ }\href@noop {} {\bibinfo {title} {quantum case-based
  reasoning (qcbr)}} (\bibinfo {year} {2021}{\natexlab{b}}),\ \Eprint
  {https://arxiv.org/abs/2104.00409} {arXiv:2104.00409 [cs.AI]} \BibitemShut
  {NoStop}%
\bibitem [{\citenamefont {Applegate}\ and\ \citenamefont
  {Cook}(1991)}]{applegate1991computational}%
  \BibitemOpen
  \bibfield  {author} {\bibinfo {author} {\bibfnamefont {D.}~\bibnamefont
  {Applegate}}\ and\ \bibinfo {author} {\bibfnamefont {W.}~\bibnamefont
  {Cook}},\ }\bibfield  {title} {\bibinfo {title} {A computational study of the
  job-shop scheduling problem},\ }\href@noop {} {\bibfield  {journal} {\bibinfo
   {journal} {ORSA Journal on computing}\ }\textbf {\bibinfo {volume} {3}},\
  \bibinfo {pages} {149} (\bibinfo {year} {1991})}\BibitemShut {NoStop}%
\bibitem [{\citenamefont {Papalitsas}\ \emph {et~al.}(2019)\citenamefont
  {Papalitsas}, \citenamefont {Andronikos}, \citenamefont {Giannakis},
  \citenamefont {Theocharopoulou},\ and\ \citenamefont
  {Fanarioti}}]{papalitsas2019qubo}%
  \BibitemOpen
  \bibfield  {author} {\bibinfo {author} {\bibfnamefont {C.}~\bibnamefont
  {Papalitsas}}, \bibinfo {author} {\bibfnamefont {T.}~\bibnamefont
  {Andronikos}}, \bibinfo {author} {\bibfnamefont {K.}~\bibnamefont
  {Giannakis}}, \bibinfo {author} {\bibfnamefont {G.}~\bibnamefont
  {Theocharopoulou}},\ and\ \bibinfo {author} {\bibfnamefont {S.}~\bibnamefont
  {Fanarioti}},\ }\bibfield  {title} {\bibinfo {title} {A qubo model for the
  traveling salesman problem with time windows},\ }\href@noop {} {\bibfield
  {journal} {\bibinfo  {journal} {Algorithms}\ }\textbf {\bibinfo {volume}
  {12}},\ \bibinfo {pages} {224} (\bibinfo {year} {2019})}\BibitemShut
  {NoStop}%
\bibitem [{\citenamefont {Boixo}\ \emph {et~al.}(2014)\citenamefont {Boixo},
  \citenamefont {R{\o}nnow}, \citenamefont {Isakov}, \citenamefont {Wang},
  \citenamefont {Wecker}, \citenamefont {Lidar}, \citenamefont {Martinis},\
  and\ \citenamefont {Troyer}}]{boixo2014evidence}%
  \BibitemOpen
  \bibfield  {author} {\bibinfo {author} {\bibfnamefont {S.}~\bibnamefont
  {Boixo}}, \bibinfo {author} {\bibfnamefont {T.~F.}\ \bibnamefont
  {R{\o}nnow}}, \bibinfo {author} {\bibfnamefont {S.~V.}\ \bibnamefont
  {Isakov}}, \bibinfo {author} {\bibfnamefont {Z.}~\bibnamefont {Wang}},
  \bibinfo {author} {\bibfnamefont {D.}~\bibnamefont {Wecker}}, \bibinfo
  {author} {\bibfnamefont {D.~A.}\ \bibnamefont {Lidar}}, \bibinfo {author}
  {\bibfnamefont {J.~M.}\ \bibnamefont {Martinis}},\ and\ \bibinfo {author}
  {\bibfnamefont {M.}~\bibnamefont {Troyer}},\ }\bibfield  {title} {\bibinfo
  {title} {Evidence for quantum annealing with more than one hundred qubits},\
  }\href@noop {} {\bibfield  {journal} {\bibinfo  {journal} {Nature physics}\
  }\textbf {\bibinfo {volume} {10}},\ \bibinfo {pages} {218} (\bibinfo {year}
  {2014})}\BibitemShut {NoStop}%
\bibitem [{\citenamefont {Preskill}(2018)}]{Preskill_2018}%
  \BibitemOpen
  \bibfield  {author} {\bibinfo {author} {\bibfnamefont {J.}~\bibnamefont
  {Preskill}},\ }\bibfield  {title} {\bibinfo {title} {Quantum computing in the
  nisq era and beyond},\ }\href {https://doi.org/10.22331/q-2018-08-06-79}
  {\bibfield  {journal} {\bibinfo  {journal} {Quantum}\ }\textbf {\bibinfo
  {volume} {2}},\ \bibinfo {pages} {79} (\bibinfo {year} {2018})}\BibitemShut
  {NoStop}%
\bibitem [{\citenamefont {De~Vogelaere}(1956)}]{de1956methods}%
  \BibitemOpen
  \bibfield  {author} {\bibinfo {author} {\bibfnamefont {R.}~\bibnamefont
  {De~Vogelaere}},\ }\bibfield  {title} {\bibinfo {title} {Methods of
  integration which preserve the contact transformation property of the
  hamilton equations},\ }\href@noop {} {\bibfield  {journal} {\bibinfo
  {journal} {Technical report (University of Notre Dame. Dept. of
  Mathematics)}\ } (\bibinfo {year} {1956})}\BibitemShut {NoStop}%
\bibitem [{\citenamefont {Marston}\ and\ \citenamefont
  {Balint-Kurti}(1989)}]{marston1989fourier}%
  \BibitemOpen
  \bibfield  {author} {\bibinfo {author} {\bibfnamefont {C.~C.}\ \bibnamefont
  {Marston}}\ and\ \bibinfo {author} {\bibfnamefont {G.~G.}\ \bibnamefont
  {Balint-Kurti}},\ }\bibfield  {title} {\bibinfo {title} {The fourier grid
  hamiltonian method for bound state eigenvalues and eigenfunctions},\
  }\href@noop {} {\bibfield  {journal} {\bibinfo  {journal} {The Journal of
  chemical physics}\ }\textbf {\bibinfo {volume} {91}},\ \bibinfo {pages}
  {3571} (\bibinfo {year} {1989})}\BibitemShut {NoStop}%
\bibitem [{\citenamefont {Shin}\ \emph {et~al.}(2014)\citenamefont {Shin},
  \citenamefont {Smith}, \citenamefont {Smolin},\ and\ \citenamefont
  {Vazirani}}]{shin2014quantum}%
  \BibitemOpen
  \bibfield  {author} {\bibinfo {author} {\bibfnamefont {S.~W.}\ \bibnamefont
  {Shin}}, \bibinfo {author} {\bibfnamefont {G.}~\bibnamefont {Smith}},
  \bibinfo {author} {\bibfnamefont {J.~A.}\ \bibnamefont {Smolin}},\ and\
  \bibinfo {author} {\bibfnamefont {U.}~\bibnamefont {Vazirani}},\ }\bibfield
  {title} {\bibinfo {title} {How" quantum" is the d-wave machine?},\
  }\href@noop {} {\bibfield  {journal} {\bibinfo  {journal} {arXiv preprint
  arXiv:1401.7087}\ } (\bibinfo {year} {2014})}\BibitemShut {NoStop}%
\bibitem [{\citenamefont {Lucas}(2014)}]{lucas2014ising}%
  \BibitemOpen
  \bibfield  {author} {\bibinfo {author} {\bibfnamefont {A.}~\bibnamefont
  {Lucas}},\ }\bibfield  {title} {\bibinfo {title} {Ising formulations of many
  np problems},\ }\href@noop {} {\bibfield  {journal} {\bibinfo  {journal}
  {Frontiers in physics}\ }\textbf {\bibinfo {volume} {2}},\ \bibinfo {pages}
  {5} (\bibinfo {year} {2014})}\BibitemShut {NoStop}%
\bibitem [{\citenamefont {Miller~C.}(1960)}]{formulacion_MTZ}%
  \BibitemOpen
  \bibfield  {author} {\bibinfo {author} {\bibfnamefont {Z.~R.}\ \bibnamefont
  {Miller~C.}, \bibfnamefont {Miller~A.}},\ }\bibfield  {title} {\bibinfo
  {title} {“integer programming formulation of traveling salesman problems.
  jacm 1960; 7 (4): 326–329. desrochers m, laporte g. improvements and
  extensions to the miller-tucker-zemlin subtour elimination constraints”},\
  }\href@noop {} {\bibfield  {journal} {\bibinfo  {journal} {Operations
  Research Letters, vol. 10, pp. 27–36}\ } (\bibinfo {year}
  {1960})}\BibitemShut {NoStop}%
\bibitem [{\citenamefont {Teplukhin}\ \emph {et~al.}(2020)\citenamefont
  {Teplukhin}, \citenamefont {Kendrick}, \citenamefont {Tretiak},\ and\
  \citenamefont {Dub}}]{teplukhin2020electronic}%
  \BibitemOpen
  \bibfield  {author} {\bibinfo {author} {\bibfnamefont {A.}~\bibnamefont
  {Teplukhin}}, \bibinfo {author} {\bibfnamefont {B.~K.}\ \bibnamefont
  {Kendrick}}, \bibinfo {author} {\bibfnamefont {S.}~\bibnamefont {Tretiak}},\
  and\ \bibinfo {author} {\bibfnamefont {P.~A.}\ \bibnamefont {Dub}},\
  }\bibfield  {title} {\bibinfo {title} {Electronic structure with direct
  diagonalization on a d-wave quantum annealer},\ }\href@noop {} {\bibfield
  {journal} {\bibinfo  {journal} {Scientific reports}\ }\textbf {\bibinfo
  {volume} {10}},\ \bibinfo {pages} {1} (\bibinfo {year} {2020})}\BibitemShut
  {NoStop}%
\bibitem [{\citenamefont {GPS}\ \emph {et~al.}(2022)\citenamefont {GPS},
  \citenamefont {Atchade-Adelomou},\ and\ \citenamefont {Bermejo}}]{GPSCODE}%
  \BibitemOpen
  \bibfield  {author} {\bibinfo {author} {\bibfnamefont {G.~A.-L.}\
  \bibnamefont {GPS}}, \bibinfo {author} {\bibfnamefont {P.}~\bibnamefont
  {Atchade-Adelomou}},\ and\ \bibinfo {author} {\bibfnamefont {S.}~\bibnamefont
  {Bermejo}},\ }\href {https://github.com/pifparfait/GPS} {\bibinfo {title}
  {Improvement in the formulation of the tsp for its generalizations type
  qubo}} (\bibinfo {year} {2022})\BibitemShut {NoStop}%
\bibitem [{\citenamefont {Jäntschi}\ \emph {et~al.}(2000)\citenamefont
  {Jäntschi}, \citenamefont {Gabriel},\ and\ \citenamefont
  {Diudea}}]{modelingmolecular}%
  \BibitemOpen
  \bibfield  {author} {\bibinfo {author} {\bibfnamefont {L.}~\bibnamefont
  {Jäntschi}}, \bibinfo {author} {\bibfnamefont {K.}~\bibnamefont {Gabriel}},\
  and\ \bibinfo {author} {\bibfnamefont {M.}~\bibnamefont {Diudea}},\
  }\bibfield  {title} {\bibinfo {title} {Modeling molecular properties by cluj
  indices},\ }\href@noop {} {\bibfield  {journal} {\bibinfo  {journal} {Match}\
  }\textbf {\bibinfo {volume} {41}} (\bibinfo {year} {2000})}\BibitemShut
  {NoStop}%
\bibitem [{\citenamefont {Zhang}\ \emph {et~al.}(2020)\citenamefont {Zhang},
  \citenamefont {Prokhorchuk},\ and\ \citenamefont {Dauwels}}]{Zhang2020}%
  \BibitemOpen
  \bibfield  {author} {\bibinfo {author} {\bibfnamefont {R.}~\bibnamefont
  {Zhang}}, \bibinfo {author} {\bibfnamefont {A.}~\bibnamefont {Prokhorchuk}},\
  and\ \bibinfo {author} {\bibfnamefont {J.}~\bibnamefont {Dauwels}},\
  }\bibfield  {title} {\bibinfo {title} {Deep reinforcement learning for
  traveling salesman problem with time windows and rejections},\ }in\ \href
  {https://doi.org/10.1109/ijcnn48605.2020.9207026} {\emph {\bibinfo
  {booktitle} {2020 International Joint Conference on Neural Networks
  ({IJCNN})}}}\ (\bibinfo  {publisher} {{IEEE}},\ \bibinfo {year}
  {2020})\BibitemShut {NoStop}%
\bibitem [{\citenamefont {Atchade-Adelomou}\ and\ \citenamefont
  {Alonso-Linaje}(2021)}]{atchade2021quantum}%
  \BibitemOpen
  \bibfield  {author} {\bibinfo {author} {\bibfnamefont {P.}~\bibnamefont
  {Atchade-Adelomou}}\ and\ \bibinfo {author} {\bibfnamefont {G.}~\bibnamefont
  {Alonso-Linaje}},\ }\bibfield  {title} {\bibinfo {title} {Quantum enhanced
  filter: Qfilter},\ }\href@noop {} {\bibfield  {journal} {\bibinfo  {journal}
  {arXiv preprint arXiv:2104.03418}\ } (\bibinfo {year} {2021})}\BibitemShut
  {NoStop}%
\end{thebibliography}%

\end{document}